\documentclass[manuscript]{emulateapj}

\usepackage{natbib}
\usepackage{lscape}

\slugcomment{To appear in the Astrophysical Journal}

\shorttitle{BLAST Survey in Vela-D}
\shortauthors{Olmi et al.}

\begin{document}

\title{The BLAST Survey of the Vela Molecular Cloud: \\
Physical Properties of the Dense Cores in Vela-D} 

\author{Luca Olmi,\altaffilmark{1,2,\dag}          %
        Peter A.~R. Ade,\altaffilmark{3}           %
	Daniel Angl\'es-Alc\'azar,\altaffilmark{1,21} %
        James J. Bock,\altaffilmark{4,5}           %
	Edward L. Chapin,\altaffilmark{6}          %
	Massimo De Luca,\altaffilmark{7}           %
        Mark J. Devlin,\altaffilmark{8}            %
        Simon Dicker,\altaffilmark{8}              %
	Davide Elia,\altaffilmark{9}               %
        Giovanni G. Fazio,\altaffilmark{10}        %
	Teresa Giannini,\altaffilmark{11}          %
        Matthew Griffin,\altaffilmark{3}           %
        Joshua O. Gundersen,\altaffilmark{12}      %
        Mark Halpern,\altaffilmark{6}              %
        Peter C. Hargrave,\altaffilmark{3}         %
        David H. Hughes,\altaffilmark{13}          %
        Jeff Klein,\altaffilmark{8}                %
	Dario Lorenzetti,\altaffilmark{11}         %
	Massimo Marengo,\altaffilmark{10}           %
        Gaelen Marsden,\altaffilmark{6}            %
        Peter G. Martin,\altaffilmark{14,15}       %
	Fabrizio Massi,\altaffilmark{2}            %
        Philip Mauskopf,\altaffilmark{3}           %
        Calvin B. Netterfield,\altaffilmark{15,16} %
        Guillaume Patanchon,\altaffilmark{17}      %
        Marie Rex,\altaffilmark{8}                 %
	Alberto Salama,\altaffilmark{18}           %
        Douglas Scott,\altaffilmark{6}             %
        Christopher Semisch,\altaffilmark{8}       %
	Howard A. Smith,\altaffilmark{10}          %
	Francesco Strafella,\altaffilmark{19}      %
	Nicholas Thomas,\altaffilmark{12}          %
        Matthew D.~P. Truch,\altaffilmark{8}       %
        Carole Tucker,\altaffilmark{3}             %
        Gregory S. Tucker,\altaffilmark{20}        %
        Marco P. Viero,\altaffilmark{15}           %
        Donald V. Wiebe\altaffilmark{16}}          %

\altaffiltext{1}{University of Puerto Rico, Rio Piedras Campus, Physics Dept., Box 23343, UPR station, San Juan, Puerto Rico}

\altaffiltext{2}{INAF, Osservatorio Astrofisico di Arcetri, Largo
E. Fermi 5, I-50125, Firenze, Italy.}

\altaffiltext{3}{Department of Physics \& Astronomy, Cardiff University, 5 The Parade, Cardiff, CF24~3AA, UK}

\altaffiltext{4}{Jet Propulsion Laboratory, Pasadena, CA 91109-8099}

\altaffiltext{5}{Observational Cosmology, MS 59-33, California Institute of Technology, Pasadena, CA 91125}

\altaffiltext{6}{Department of Physics \& Astronomy, University of
British Columbia, 6224 Agricultural Road, Vancouver, BC V6T~1Z1,
Canada}

\altaffiltext{7}{LERMA-LRA, CNRS UMR8112, Observatoire de Paris and Ecole
Normale Sup\'erieure, 24 Rue Lhomond, 75231 Paris cedex 05, France}

\altaffiltext{8}{Department of Physics and Astronomy, University of Pennsylvania, 209 South 33rd Street, 
Philadelphia, PA 19104}

\altaffiltext{9}{Universidade de Lisboa - Faculdade de Ciencias,
Centro de Astronomia e Astrof\'isica da Universidade de Lisboa,
Obs\'ervatorio Astron\'omico de Lisboa,
Tapada da Ajuda, 1349-018 Lisboa, Portugal}

\altaffiltext{10}{Harvard-Smithsonian Center for Astrophysics, Cambridge, MA 02138}

\altaffiltext{11}{INAF, Osservatorio Astronomico di Roma, Via Frascati 33, I-00040
Monteporzio Catone, Roma, Italy.}

\altaffiltext{12}{Department of Physics, University of Miami, 1320 Campo Sano Drive, Carol Gables,
FL 33146}

\altaffiltext{13}{Instituto Nacional de Astrof\'isica \'Optica y Electr\'onica (INAOE), 
Aptdo. Postal 51 y 72000 Puebla, Mexico}

\altaffiltext{14}{Canadian Institute for Theoretical Astrophysics, University of Toronto, 60 
St. George Street, Toronto, ON M5S~3H8, Canada}

\altaffiltext{15}{Department of Astronomy \& Astrophysics, University of Toronto, 50 St. 
George Street, Toronto, ON  M5S~3H4, Canada}

\altaffiltext{16}{Department of Physics, University of Toronto, 60 St. George Street, Toronto, ON M5S~1A7, Canada}

\altaffiltext{17}{Laboratoire APC, 10, rue Alice Domon et L{\'e}onie Duquet 75205 Paris, France}

\altaffiltext{18}{European Space Astronomy Centre, Villanueva de la Canada, Apartado 78, 28691 Madrid, Spain}

\altaffiltext{19}{Dipartimento di Fisica, Univ. del Salento, CP 193,
I-73100 Lecce, Italy.}

\altaffiltext{20}{Department of Physics, Brown University, 182 Hope Street, Providence, RI 02912}

\altaffiltext{21}{Department of Physics, University of Arizona, 1118 E. 4th Street, Tucson, AZ 85721}

\altaffiltext{\dag}{\url{olmi.luca@gmail.com, olmi@arcetri.astro.it}}

\begin{abstract}

The Balloon-borne Large-Aperture Submillimeter Telescope (BLAST) carried out a 250, 350 and 
500\,\micron\ survey of the galactic plane encompassing the Vela Molecular Ridge, 
with the primary goal of identifying 
the coldest dense cores possibly associated with the earliest stages of star formation.  
Here we present the results from observations of the Vela-D region, covering about $4\,{\rm deg}^2$, 
in which we find 141 BLAST cores.  We exploit existing data taken with the {\it Spitzer\/} MIPS, 
IRAC and SEST-SIMBA instruments to constrain their (single-temperature) spectral energy distributions, 
assuming a dust emissivity index $\beta=2.0$.  This combination of data allows us to determine the 
temperature, luminosity and mass of each BLAST core, and also enables us to separate starless 
from proto-stellar sources. We also analyze the effects that the uncertainties on the 
derived physical parameters of the individual sources have 
on the overall physical properties of starless and
proto-stellar cores, and we find that there appear to be a smooth transition from
the pre- to the proto-stellar phase.  
In particular, for proto-stellar cores we find a correlation between
the MIPS24 flux, associated with the central protostar, and the temperature of the dust
envelope.  We also find that the core mass function of the Vela-D cores has a slope consistent 
with other similar (sub)millimeter surveys.  

\end{abstract}


\section{INTRODUCTION }


Stars form in dense, dusty cores of molecular clouds, but little is known
about their origin, their evolution, and their detailed physical properties.
In particular, the relationship between the core mass function (CMF) 
and the stellar initial mass function (IMF) is poorly understood
\citep{mckee2007}.  One of the reasons for this lack of understanding,
from the observational
point of view, is the difficulty in selecting a statistically significant
sample of truly {\it pre-stellar} cores from an otherwise unremarkable collection
of high column density features.

Pre-stellar cores represent a very early stage of the star formation (SF)
process, before collapse results in the formation of a central protostar. 
The physical properties of these cores can reveal important clues about their
 nature; mass, spatial distributions, and lifetime are important
diagnostics of the main physical processes leading to the formation of
the cores from the parent molecular cloud.
In addition, a comparison of the CMF to the IMF
may help to understand what processes are responsible for
further fragmentation of the cores, and thus the
determination of stellar masses. Therefore, large samples of {\it bona-fide}
pre-stellar cores are important for comparison of observations
with various SF models and scenarios.


Two major technological advances have revolutionized the study of the
 mass distribution in molecular clouds and the characterization of the
earliest phases of SF: the Infrared Astronomical Satellite ({\it IRAS}\,) 
satellite, and the advent of submillimeter bolometer arrays
such as MAMBO, SIMBA, and SCUBA. 
(Sub)millimeter surveys have been vitally important
for probing the nature of SF,
and in particular have represented the only way of deriving
statistically-significant samples of pre- and proto-stellar objects.
However, these instruments have probed the Rayleigh-Jeans tail of the spectral energy distribution 
(SED) of these cold objects, far from its peak. Therefore, these surveys have been 
limited by their relative inability to measure the temperature 
(e.g., \citealp{motte1998}), producing large
uncertainties in the derived luminosities and masses.  Recent surveys
with the MIPS instrument of the {\it Spitzer Space Telescope} 
are able to constrain the temperatures of warmer
objects \citep{car05}, but the youngest and coldest objects are
potentially not detected, even in the long-wavelength {\it Spitzer}
bands.  Other ground-based surveys conducted at wavelengths
$\la 450 \,$\micron\ have been affected by low sensitivity due to
atmospheric conditions  at short submillimeter wavelengths (e.g., \citealp{kirk2005, wu2007}). 

The Balloon-borne Large-Aperture
  Submillimeter Telescope, BLAST  \citep{pascale2008}, 
carried out two long duration balloon  science flights. During
the first (BLAST05), BLAST mapped several galactic star
forming regions \citep{chapin2008, truch2008}.
During the second science flight in 2006, BLAST06
mapped about  50\,deg$^2$ in the Vela Molecular Ridge (VMR) \citep{netterfield2009}.
Until the more extensive results from {\it Herschel} become available, BLAST
is unique in its ability to 
detect and characterize cold dust emission from both starless and proto-stellar 
sources, constraining the temperatures of objects with $T \la 25$\,K
using its three-band photometry (250, 350, and 500\,\micron) near the peak of the
cold core SED.  

The VMR is a giant molecular cloud complex within the
galactic plane, in the area
$260 \degr \la l \la 272 \degr$ and $-2 \degr \la b \la 3 \degr$,
hence located outside the solar circle. It
was first studied in detail by \citet{mur:may}, using
millimeter low-resolution observations in the CO(1--0) transition.
They subdivided the VMR
into four main regions, named A, B, C, and D.\@ \citet{Lis92}
further discussed the issue of distance, finding that clouds A, C, and D are
at $700 \pm 200$ pc, whereas B seems to be at $\sim 2000$ pc.
In this work we have selected the Vela-D region because of the
large amount of ancillary data available, both
in the continuum, particularly in the mid- (MIR) and far-infrared (FIR),
and also in terms of spectral line observations. 

\citet{Lis92} and \citet{lor93} identified a
number of VMR sites that host low- and intermediate-mass (proto)stars whose
infrared SEDs are consistent with those of Class I sources. Later,
\citet{massi99, massi00, massi03} analyzed NIR images 
of a sample of those sites, finding that
the {\it IRAS} sources with $L_{\rm bol} > 10^{3} \,$L$_{\sun}$ are associated with
small embedded young stellar clusters. Their IMF appears
consistent with that of field stars and
their age is of the order of $10^{6}$ yrs \citep{massi06}.
Remarkably, cloud D exhibits a lack of massive (O-type) stars with respect
to what is expected from a standard IMF \citep{massi06}.

\begin{figure}
\centering
\includegraphics[width=\linewidth]{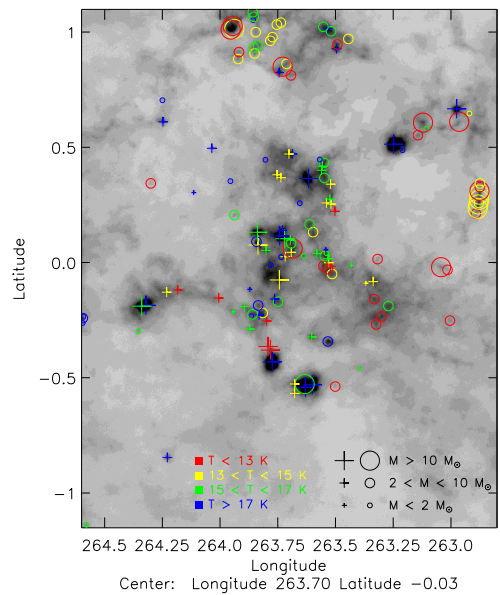}
\caption{
The gray-scale image shows the BLAST 250\,\micron\ map of Vela-D, with galactic coordinates
in degrees. Superimposed are the locations of both starless (open circles) and proto-stellar
(crosses) cores (see \S\ref{sec:phot} and \S\ref{sec:comparison}). 
 The size of both crosses and circles indicates the mass range 
of each core  and color-coding indicates the core temperature (see legend).
See \S\ref{sec:masslum} for a discussion of the cores' physical parameters.
 }
\label{fig:blast250}
\end{figure}

New millimeter observations, both spectral line \citep{elia2007}  and continuum \citep{massi07},
have shown that all the {\it IRAS} sources associated 
with embedded young clusters are also
associated with one or more dense molecular cores, confirming the inferred
young age of the clusters. On-going SF activity in Vela-D  is also confirmed by the presence
of collimated jets originating from objects embedded in some of the clusters
\citep{lore02, delu07}.
A first census of young stellar objects (YSOs)
 throughout the area of cloud D mapped by \citet{massi07} and 
\citet{elia2007}, was carried out by \citet{gianni07}
using {\it Spitzer} MIPS observations at 24 and $70\,$\micron\  
(hereafter indicated as MIPS24 and MIPS70, respectively). Most of the millimeter cores
identified by \citet{massi07} were found to be associated with red,
cold objects. Cloud~D appears therefore as an active star-forming region with
both distributed and clustered SF in progress, hosting a large sample
of young objects in different evolutionary stages.

In this paper we use both
BLAST and archival data to determine the  SEDs, and thus physical parameters, 
of each source detected by BLAST.
In \S\ref{sec:obs} we describe the BLAST observations and also the archival data
that have been used in this work.  In \S\ref{sec:phot} we describe the identification
of millimeter,  MIR and FIR counterparts of the BLAST cores, and also
describe the multi-band photometry. This photometry is then used to construct
the SED of each source in \S\ref{sec:sed}, which allow us to derive
the physical parameters for starless and proto-stellar cores, discussed in \S\ref{sec:comparison}.
We finally draw our conclusions in \S\ref{sec:concl}.

\section{OBSERVATIONS }
\label{sec:obs}

\subsection{BLAST observations } 
\label{sec:blastobs}

BLAST06 is a 1.8\,m Cassegrain telescope, whose under-illuminated primary
mirror has produced in-flight beams with FWHM of
36\,\arcsec, 42\,\arcsec, and 60\,\arcsec\ at 250, 350, and 500\,\micron,
respectively.  The camera consists of three silicon-nitride ``spider
web'' bolometer arrays \citep{turner2001} almost identical to those
for SPIRE on {\it Herschel} \citep{griffin2008}, with 149, 88, and 43
detectors at 250, 350, and 500\,\micron, respectively, organized in a hexagonal
close-packed pattern. 
The raw BLAST data are reduced using a common pipeline. More details can be found in
the BLAST05 papers \citet{pascale2008},  
 \citet{patanchon2008}, and \citet{truch2008}. 
The absolute gain of the instrument (including antenna efficiency), determined 
from regular observations of the evolved star VY\,CMa, is measured with highly correlated absolute 
uncertainties of 10\%, 12\%, and 13\% at 250, 350, and 500\,\micron, 
respectively \citep{truch2009}. 
Finally, maximum likelihood maps and noise estimates 
are made (\citealp{patanchon2008}, \citealp{wiebe2008}).

\begin{figure}
\centering
\includegraphics[width=\linewidth]{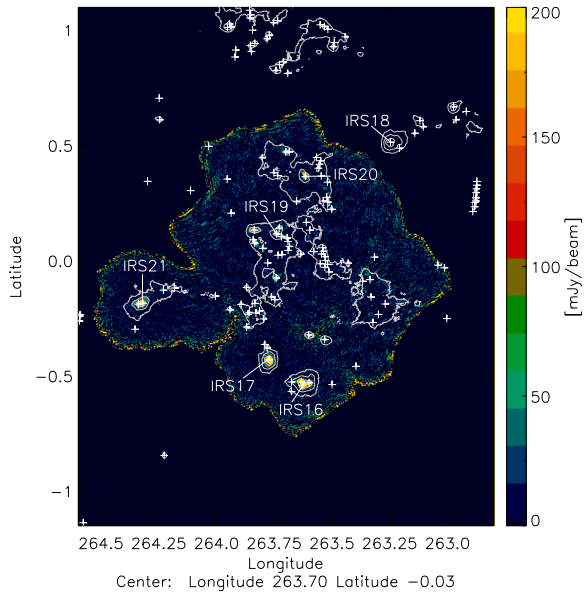}
\caption{
SIMBA 1.2\,mm emission from \citet{massi07}.
Overplotted are the BLAST 250\,\micron\  contours, and the positions of the
compact BLAST sources (``+'' signs). Also shown are the positions of the main bright infrared
sources of \citet{Lis92}.
Galactic coordinates are in degrees.
  }
\label{fig:blast_simba}
\end{figure}

The deepest map of the Galactic Plane obtained by BLAST06 spans $\sim 10$\,deg in 
Galactic longitude and $\sim 5$\,deg in Galactic latitude in the constellation Vela 
(including the VMR) which we will call the ``Vela deep'' map hereafter. The Vela deep map
is part of a much larger ($\sim 200$\,deg$^2$) but shallower map, called the ``Vela wide''
map\footnote{http://www.blastexperiment.info/}, which is still unpublished.
As noted in the introduction, we will concentrate here on the Vela-D
region, which for practical reasons we define as the area of the Vela deep map
contained within $262\fdg80 < l < 264\fdg60$ and  
$-1\fdg15 < b < 1\fdg10$. In this area we find a total of 141 sources
from the BLAST catalog of Vela deep, shown in Figure~\ref{fig:blast250} 
 and described later in \S\ref{sec:phot} and \S\ref{sec:comparison}.
The high signal-to-noise ratio of the BLAST maps allows us to infer 
intrinsic source sizes even below the beam FWHM scale by deconvolving the BLAST beam from the
measured source FWHM (using a Gaussian source surface brightness profile: see \S\ref{sec:blastflux}
and \citealp{netterfield2009}).  
We find that the sizes from the fit are broader than the
intrinsic beam size, with a typical diameter of 62\,\arcsec\  (measured at 250\,\micron), which corresponds to an
intrinsic deconvolved source size of $\simeq 0.15\,$pc at the distance of the VMR.

\subsection{Archival data }
\label{sec:archive}

\subsubsection{SEST data}
\label{sec:archiveSIMBA}

Figure~\ref{fig:blast_simba} shows the
$1 \times 1$ deg$^{2}$ area of Cloud~D mapped with SIMBA at the
SEST (ESO, La Silla, Chile) in the 1.2-mm continuum (\citealp{massi07}). 
At this wavelength, the SIMBA beam is $\simeq 24\,\arcsec$, i.e. $\sim 0.08$\,pc at 700\,pc.
The r.m.s. in the map is in the range $14-40$\,mJy\,beam$^{-1}$, 
but it is $\sim 20$\,mJy\,beam$^{-1}$ over most
of the map \citep{massi07}.
By using the algorithm {\small CLUMPFIND} \citep{williams1994},
the emission was subdivided by \citet{massi07} into
29 dense molecular cores, with deconvolved sizes in the range $0.03$--$0.25$
pc and masses in the range $0.4$--88\, ${\rm M}_{\sun}$. \citet{massi07} discarded a total 
of 26 additional cores because they were smaller than the SEST beam. However, 
\citet{delu07} showed that most of these objects are probably real millimeter
sources associated with SF and not
just instrumental artifacts. The two sets of objects and the assumptions made
to derive their physical properties are discussed in \citet{massi07}.
Their Table~1 lists the {\it robust} sample of 29 sources (named ``MMS'' by the authors), while
their Table~2 lists the sample of possible cores (named ``umms''). Our own analysis
of the SIMBA data and association with BLAST sources are discussed in \S\ref{sec:simba}.

\begin{figure}
\centering
\includegraphics[width=\linewidth]{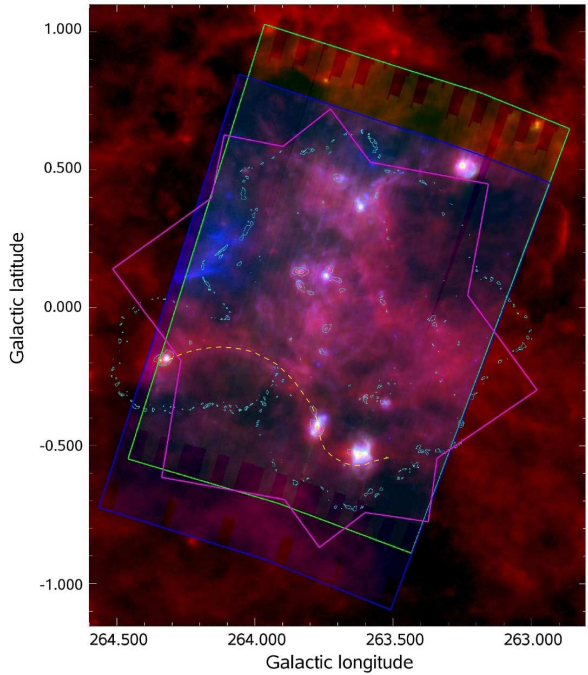}
\caption{
False color image of the Vela-D map using MIPS24  
for blue, MIPS70  for green, and BLAST 250\,\micron\  
for red. Color in this image is an indicator of temperature, with blue regions
usually being warmer and red regions being cooler.
The background image is the BLAST 250\,\micron\ map, the blue-shaded region represents
the MIPS24 map, the region outlined by the green contour is the map covered by MIPS70,
the irregular purple line outlines show the region covered by IRAC and the contours represent the
SIMBA map.
The ``{\it S}''-shaped dashed, yellow contour shows the approximate location of the filament
containing sources IRS16, IRS17 and IRS21 (see \S\ref{sec:morph} and Figure~\ref{fig:blast_simba}).
  }
\label{fig:map_coverage}
\end{figure}

\subsubsection{IRAC, MIPS and  {\it Akari}}

Concerning the {\it Spitzer} MIPS data \citep{rieke2004},
in this paper we make use of maps (covering $\sim 1.5$~deg$^2$, see Figure~\ref{fig:map_coverage}) 
at 24 and 70\,\micron\  of Vela-D 
which have already been published \citep{gianni07},
and the reader is referred to that paper where the image mosaicking,
artifact removal and details relative to
source finding and flux extraction (see also \S\ref{sec:MIPSphotom})
are discussed.

\begin{figure}
\centering
\includegraphics[width=\linewidth]{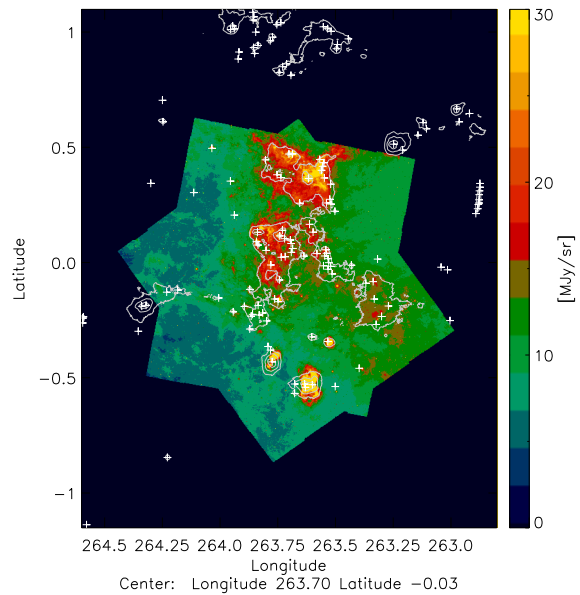}
\caption{
IRAC  emission at 8\,\micron\
overplotted with the BLAST 250\,\micron\  contours and the positions of 
compact BLAST sources (``+'' signs).
We note that some of the BLAST sources
occur at minima, or darker regions, in the diffuse 8\,\micron\  emission,
possibly indicating that the submillimeter objects are (associated with)
infrared dark clouds in the foreground (see also \S\ref{sec:mips24irac}).
Galactic coordinates are in degrees.
  }
\label{fig:irac8_blast}
\end{figure}

To characterize the BLAST sources at the shortest possible wavelengths
we also use  {\it Spitzer} IRAC \citep{fazio2004}
observations, in particular those in the 8\,\micron\ band, covering part of
Vela-D (see Figures~\ref{fig:map_coverage} and \ref{fig:irac8_blast}). 
The sensitivity achieved at 8\,\micron\ was about
50$\,\mu$Jy. %
Moreover, because {\it Spitzer} is diffraction limited longward of
 $\sim$10\,\micron, the spatial resolution in the MIPS
bands is $\simeq 6\,\arcsec$ at 24\,\micron, and $\simeq 18\,\arcsec$ at 70\,\micron.
We note that the present paper is based on a sub-sample of a much larger catalog
incorporating both IRAC and MIPS data \citep{stra2009}.

The {\it Akari}/FIS survey at 60, 90, 140 and 160\,\micron\  \citep{yama2008}
is useful because of its much higher sensitivity compared
to {\it IRAS}, and also because of the two longer wavelength bands, making it more sensitive
to cold dust emission. 
An early {\it Akari} catalog has been used to find FIR counterparts to the BLAST
sources, but without the actual maps we have not been able to 
directly compare the brightness distribution with that observed at both longer and shorter
wavelengths.

\subsection{Morphology}
\label{sec:morph}

Figure~\ref{fig:map_coverage} shows a false color image of the Vela-D map, and
the 1.2-mm SIMBA map of the Vela-D region \citep{massi07}
is shown in Figure~\ref{fig:blast_simba}. 
The brightest regions in Figure~\ref{fig:blast_simba} are clearly associated with known
SF regions, including IRS16, IRS17, IRS18, IRS19, IRS20 and IRS21 
(following the designation adopted by \citealp{Lis92}).
IRS16 is located at the center of the \ion{H}{2}
region 263.619-0.533 \citep{caswell1987}, and the
most populated areas of the map with BLAST cores are clearly
the regions around sources IRS19 and IRS20, which show a
complex structure composed of filaments, diffuse emission and
clusters of dense cores.

The nature of the IRS sources has been analyzed in detail by
\citet{massi03}, who concluded that most of these objects
are probably precursors of intermediate mass stars ($M\sim 2 - 10\, {\rm M}_{\odot}$)
which may still be in an accretion stage (i.e., Class I sources), thus
explaining their red colors. 
 By comparing Figures~\ref{fig:blast250} to \ref{fig:map_coverage}
one can also see that IRS16, IRS17 and IRS21 are all
part of a long, ``{\it S}''-shaped filament containing several BLAST and
SIMBA cores, including the most massive SIMBA core, MMS4.
The presence and nature of this, as well as other arc-like features in 
the Vela-D region, has been analyzed by \citet{elia2007}.

Figure~\ref{fig:irac8_blast} shows an overlay of the IRAC 8\,\micron\ (hereafter IRAC8) image 
with the BLAST 250\,\micron\ contours. Although both trace neutral material, 
the 8\,\micron\ emission is known to be heavily influenced by PAH emission, and 
requires UV illumination, instead of being a simple tracer of column density.
The IRAC8 map can also be used to check for extinction effects.
In Figure~\ref{fig:irac8_blast} the BLAST contours follow the general 
distribution of the 8\,\micron\ emission. However, there are a number
of BLAST compact sources, along with the neutral material associated with them,
that correspond in position with local minima or darker regions in the diffuse 8\,\micron\ emission.
This effect has also been observed in Vulpecula by \citet{chapin2008}.
However, analysing whether this is the result of chance coincidence or rather 
it is indeed the BLAST cores that are producing the so-called infrared dark clouds (IRDCs; e.g.,
\citealp{egan1998, simon2006}) was beyond the scope of this work.

Finally, we note that the BLAST cores are unlikely to be extragalactic bright sources. Using
the number counts in \citet{patanchon2009} the expected number of extragalactic sources
at 250\,\micron\ with flux density greater than 0.3\,Jy is $\simeq 1\,$deg$^{-2}$, while 
at the level of a few Jy (the observed minimum source flux density at 250\,\micron), this
number is $<< 1\,$deg$^{-2}$.

\section{SOURCE PHOTOMETRY AND ASSOCIATIONS WITH COMPACT SOURCES}
\label{sec:phot}

\subsection{Source and Flux extraction in the BLAST maps}
\label{sec:blastflux}

The source extraction and brightness estimation techniques applied to the BLAST maps 
here are similar to the method used during analysis 
of the BLAST05 data \citep{chapin2008}.
Candidate sources are identified by finding peaks after a Mexican Hat Wavelet 
type convolution (see, e.g., \citealp{barnard2004}).  
The candidate lists from 250 and 350\,\micron\  are then merged and fluxes at 
all three bands extracted by fitting a compact Gaussian profile to the source. 
Sources are not identified at 500\,\micron\  due to the greater source-source and 
source-background confusion resulting from the lower resolution. 
 However, a Gaussian is refitted also in the 500\,\micron\ map using the size 
and location parameters determined at the shorter wavelengths (the size of the 
Gaussian is convolved to account for the differing beam sizes).
Using this technique, more than one thousand compact cores were identified in the Vela deep map 
\citep{netterfield2009}. 

The catalog of BLAST source positions in Vela-D, including 
flux densities, is given in Table~\ref{tab:src}. This list is taken
as the reference catalog for all subsequent associations with compact sources
detected at all other wavebands described below.
The cross-correlation between BLAST sources and objects listed in other catalogs
is performed using positional criteria. An archive object is
considered to be associated with a BLAST source when their separation is less
than the following search radius:
\begin{equation}
R_{\rm search} = [D_{\rm BLAST}^2 + (\epsilon_{\rm ptg}^{\rm archive})^2]^{1/2},
\end{equation}
where $\epsilon_{\rm ptg}^{\rm archive}$ is the pointing error associated with the position 
of the archive source, and we have defined
\begin{equation}
D_{\rm BLAST} = [(D_{\rm dec}/2)^2 + (\epsilon_{\rm ptg}^{\rm BLAST})^2 + 
\epsilon_{\rm extr}^2]^{1/2}, \\
\end{equation}
where $D_{\rm dec}$ is the deconvolved FWHM of the BLAST source. Both 
$\epsilon_{\rm ptg}^{\rm BLAST}$,  the BLAST pointing error, 
and $\epsilon_{\rm extr}$,  the uncertainty in the source position 
due to the source finding technique, have been estimated to be  $\sim 5$\,\arcsec.

\subsection{The millimeter continuum: SIMBA data }
\label{sec:simba}

Due to the smaller area and lower signal-to-noise ratio of the SIMBA map (\S\ref{sec:archiveSIMBA}),
when performing a cross-correlation between the BLAST and SIMBA sources
we find a SIMBA counterpart to only 31 BLAST sources, out of 141 in our selected
area of Vela-D.    
We determine the positional offset from each candidate SIMBA counterpart to the nearest 
BLAST source position 
and find that the average positional offsets between the BLAST and SIMBA positions 
are $-3.5$\,\arcsec\ and $-1.4$\,\arcsec\ in galactic longitude
and latitude, respectively, whereas the standard deviations in these two
axes are 7.2\,\arcsec\ and 6.0\,\arcsec. The offsets are also randomly distributed,
with no particular recognizable pattern, 
and we attribute them to instrumental pointing and to the different
algorithm used for source finding in the BLAST and SIMBA maps.

The cross-correlation between the BLAST and SIMBA catalogs follows the procedure outlined
in \S\ref{sec:blastflux}, assuming $\epsilon_{\rm ptg}^{\rm SIMBA} \simeq 5$\,\arcsec\ \citep{massi07}.
We have used the flux densities listed by \citet{massi07} for the MMS sources, whereas
for the less robust catalog of umms sources (\S\ref{sec:archiveSIMBA}) we have performed direct
aperture photometry in the SIMBA map.  However, if the umms source was found near to
another SIMBA source, we retained the original catalog flux density value.
In the case that more than one SIMBA source is positionally associated with a single BLAST object,
or it is sufficiently close to the BLAST source to affect the photometry,
the flux densities of all associated sources are added and used as an upper limit in the BLAST source SED.
For those BLAST sources without a SIMBA counterpart (MMS or umms), we
estimate an upper limit to the flux density by performing aperture photometry
centered over the BLAST coordinates.

\subsection{The MIR and FIR}

\subsubsection{{\it IRAS} and {\it Akari}}
\label{sec:iras}

In Vela-D we find that several BLAST sources have counterparts in
the {\it IRAS} Point Source Catalogue version 2.0 \citep[PSC,][]{helou1988}. 
For the cross-correlation between the {\it IRAS}-PSC and the BLAST catalogs 
we set the positional error $\epsilon_{\rm ptg}^{IRAS}$ 
equal to the semi-major axis of the {\it IRAS} error ellipse. 
Identifications from the {\it IRAS}-PSC were found for 26 of the 141 BLAST 
sources. 
For sources that lack PSC counterparts or measurements in any of the
{\it IRAS} bands we produce measurements or upper limits directly from
the {\it IRAS}\, Galaxy Atlas maps \citep[IGA,][at 60 and 100\,\micron, 
hereafter IRAS60 and IRAS100]{cao1997}.
The point spread functions (PSFs) vary across these maps, showing strong
elongation along the scan direction. 
We have used aperture photometry to measure flux densities in the IGA maps. However,
the relatively poor angular resolution of {\it IRAS} prevents unambiguous determination of
the flux in the same volume as was probed by the BLAST sources, due to both
extended halo emission and source crowding.  Consequently, the {\it IRAS}-PSC fluxes
are used as upper limits only.

We have found the {\it Akari} counterparts associated with all BLAST sources in Vela-D.
However, the {\it Akari} fluxes have not been used in estimating the best-fit SED
(\S\ref{sec:sed}), although they are actually shown on the source SED in the two examples presented in
Figures~\ref{fig:SEDex1} and \ref{fig:SEDex2}. In fact, we have noted that the
{\it Akari} fluxes from the catalog available to us (which does not include the latest
calibration) 
very often do not quite fall on the best-fit SED determined using all other
wavebands. For this reason we have conservatively determined the SED and the physical
parameters of the BLAST sources from the SIMBA, BLAST and MIPS70 fluxes only.

\subsubsection{IRAC and MIPS photometry }
\label{sec:MIPSphotom}

The IRAC maps were analyzed by means of the DAOPHOT photometric package
(see \citealp{stra2009}).  Despite the
marginal sampling of the PSF by the IRAC cameras, particularly in the
3.5\,\micron\ and 4.6\,\micron\ bands, the PSF photometry
was preferred because of its better accuracy, with respect to aperture photometry,
in dealing with both crowded regions and variable backgrounds. After selection of
a sample of bright and unsaturated sources, a point response 
function\footnote{For a comparison between 
the PRF and PSF see for example 
http://ssc.spitzer.caltech.edu/irac/psf.html.}
(PRF) was derived in each band and used to fit the sources detected at
S/N $>$ 5 to obtain the corresponding photometry \citep{stra2009}. 
For the purpose of cross-correlation between the BLAST and IRAC catalogs, 
we assumed $\epsilon_{\rm ptg}^{\rm IRAC} \simeq 2$\,\arcsec.

As for IRAC, the {\it Spitzer} MIPS observations were
exploited to search for possible counterparts to the BLAST sources, 
also assuming $\epsilon_{\rm ptg}^{\rm MIPS} \simeq 2$\,\arcsec.
Because the MIPS mosaics of the Vela-D cloud are characterized by a particularly
strong background level and gradient, we adopted an appropriate spatial filter
to smooth all the spatial scales with sizes larger than 20 pixel
(corresponding to 50\,\arcsec\ and 80\,\arcsec\ at 24 and 70\,\micron, respectively), preserving the
small scale fluctuations. Then, as for  IRAC, we performed PSF photometry.
This approach allowed us to enlarge the original sample discussed in \citet{gianni07}.

Because the MIPS70 data lie on the Wien side of the SED, they can
strongly constrain the SED (\S\ref{sec:sed}), and it is thus important to determine whether the 
integrated emission in each waveband comes from the same volume of material. 
Submillimeter dust cores, particularly proto-stellar
cores, are likely to have density as well as temperature gradients, implying that the bulk of 
the emission at different wavelengths may come from substantially different volumes
of gas and dust, becoming progressively more optically thick at shorter wavelengths. 
The power-law density (and, possibly, temperature) profiles of the cores is then 
coupled with the instrumental response at each waveband, making the exact definition
of the ``same'' core at different wavelengths an imprecise concept at best.
However, we note that in a typical proto-stellar core the wavelength of a photon carrying the 
mean energy of the SED is $\simeq 50\,$\micron\ and the radius where this photon escapes
the cloud is estimated to be $\simeq 13 \,$AU (e.g., \citealp{sp2004}).
Thus, we think that the MIPS70 emission can still be mostly associated with the envelope
of the proto-stellar core.

We investigated alternative photometric techniques and, in particular, 
we compared aperture photometry and a method similar to that used in the BLAST maps
(\S\ref{sec:blastflux}),  which may be of relevance to candidate flux-extraction
techniques to be used in the forthcoming {\it Herschel} multi-wavelength maps. 
We first convolve the MIPS70 map to the BLAST250 resolution,
and then we fit each compact source in the resulting map with
the {\it same} Gaussian profile previously extracted from the BLAST maps.
However, source crowding may seriously affect the accuracy of aperture photometry, 
while intrinsically different brightness distributions in the BLAST and MIPS70 wavebands
make the extracted flux from fixed-position Gaussians quite unreliable.
We thus decided to use catalog values, as determined
from PSF photometry,  except in three (isolated) sources where the catalog values were clearly
incosistent with the SED constrained by SIMBA and BLAST data points. In these few cases
aperture photometry
was used\footnote{Following the MIPS manual at http://ssc.spitzer.caltech.edu/mips/apercorr/
we used an aperture radius
$R_{\rm apr}=8.75$~pixel (with 1~pixel=4\,\arcsec\ in the MIPS70 image)
and an annulus region delimited by two circles of radii $R_{\rm sky1}  = 9.75$~pixel and
$R_{\rm sky2}  = 16.25$~pixel, and an additional correction factor $c = 1.308$.}.
For those BLAST sources without a MIPS70 counterpart, an upper limit to the emission was determined
also using aperture photometry.

We also checked the MIPS maps for the possible occurrence of detector saturation. 
In accordance with the MIPS manual\footnote{http://ssc.spitzer.caltech.edu/documents/SOM/} 
we verified that the elemental integration time,
which represents the ultimate threshold to estimate a possible saturation,
was short enough that none of the detected sources are saturated at 70\,\micron.

\section{MILLIMETER-MIR SEDs}
\label{sec:sed}

\subsection{The Approach}

In this section we describe how,  using
the SIMBA, BLAST and MIPS70 photometry (i.e., excluding {\it Akari} data points, see \S\ref{sec:iras}) 
discussed in the previous sections, 
we fit a SED for each BLAST core. 
However, this is complicated by the possibility that some
BLAST cores may be composed by multiple sources. In fact, 
although BLAST achieved an unprecedented angular resolution at 250\,\micron,
this is partially compensated by the relatively large distance of Vela-D, 
thus resulting in a spatial resolution of
about 0.1\,pc. This scale is comparable to the average dense core diameters
found by \citet{motte2007} in Cyg-X and by \citet{rathborne2008} in
the Pipe nebula. The 0.1\,pc scale is also comparable with the average diameter of ammonia cores
found by \citet{jijina1999}, although it is considerably larger than the 
diameter found for low-mass cores in nearby SF regions by \citet{ward1999} and \citet{enoch2008}. 
However, we also find that of the 31 SIMBA sources associated with
BLAST cores, only 4 are actually resolved as a double-core system 
by the smaller beam of the SEST telescope,
suggesting that either the typical scale-length of the cores in Vela-D is considerably smaller
than $\sim 0.1$~pc, or there are relatively few multiple sources.

%
%
 \begin{figure}
 \centering
 \includegraphics[width=3in]{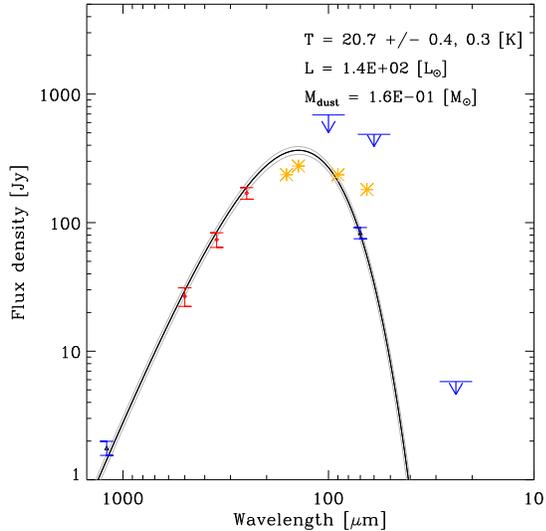}
 \includegraphics[width=3in]{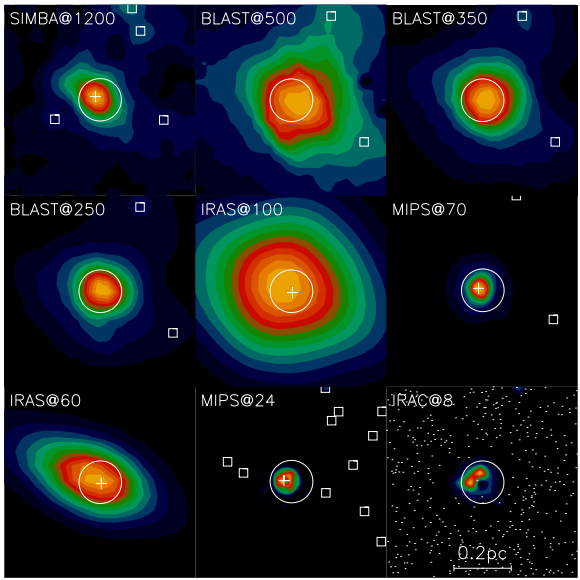}
 \caption{
{\it Left.} SED of one of the brightest sources in Vela-D, BLAST J084848-433225.
Various upper limits are shown and the asterisks represent the {\it Akari} fluxes
({\it not} used in the fit to the SED).  The black line shows the best-fit modified blackbody,
at wavelengths $> 24$\,\micron, whereas gray lines show the 68\% confidence envelope
of modified blackbody models from Monte Carlo simulations in which the dust emissivity index
is fixed at $\beta=2.0$.  {\it Right.} Maps of this source at various wavelengths,
from left to right and from top to bottom: SIMBA, BLAST500, BLAST350,
BLAST250, IRAS100, MIPS70, IRAS60, MIPS24 and IRAC8. Each map covers a
$200\,\arcsec \times 200\,\arcsec$ region. The large open circle  represents the
observed FWHM of the BLAST core.
The ``+'' sign marks the position of each specific catalog
source (except IRAC) falling within the search radius (\S\ref{sec:blastflux}) from the
nominal position of the BLAST core. 
In the IRAC map, catalog sources are represented by the white dots.
The open squares represent other catalog sources falling outside the search radius.
   }
\label{fig:SEDex1}
\end{figure}

In addition, we note that for the typical mass of the cores in Vela-D (a few solar masses,
see \S\ref{sec:masslum}) the power-law relation $M \propto R^{2.6}$ found by 
\citet{lada2008} between the core mass and the core radius, would predict
(if extrapolated to masses $M \ga 1 \, {\rm M}_{\odot}$) a core radius of about 0.05~pc,
well matched with the linear resolution of BLAST at the distance of Vela-D.
Furthermore, based on data from \citet{beltran2006}, who surveyed 
more massive cores, one can also find that $M \propto R^p$ ($p\simeq 1.6$, Cesaroni, priv. comm.), 
quite similar to the mass-size relation found by \citet{massi07}, $M \propto R^{1.7}$, 
in Vela-D itself.  These findings indicate that for core masses larger than about 
$1 {\rm M}_{\odot}$ almost all sources have diameters larger than about $0.05-0.1\,$pc.

Nevertheless, even in the BLAST cores that are indeed individual entities, 
the source would still likely be 
composed of regions at different temperatures, typically a
warmer core (with or without a protostar or proto-stellar cluster) 
embedded in a colder and less dense medium (except for those cores possibly heated 
from an external source).  The variations in 
physical parameters such as temperature and density are likely to be
much less critical in starless cores (e.g., \citealp{andre1999}).
Our goal is thus to use a simple, single-temperature SED model to fit the 
sparsely sampled photometry described in \S\ref{sec:phot}, 
which will allow us to infer the main physical
parameters of each core: mass, temperature and luminosity. These
quantities must be interpreted as a parameterization of a more 
complex distribution of temperature and density in the core and the
equally complex response of each instrument to these physical conditions.

\subsection{SED fitting with modified blackbody function}
\label{sec:SEDfit}

Following the method described by \citet{chapin2008}
we assume optically-thin emission from an isothermal modified
blackbody,
\begin{equation}
  S_{\nu} = A
  \left(\frac{\nu}{\nu_0}\right)^{\beta} B_{\nu}(T),
\label{eq:sed}
\end{equation}
where $A$ is a constant, $B_{\nu}(T)$ is the Planck
function, $\beta$ is the dust emissivity index, and the emissivity
factor  is normalized at a fixed frequency $\nu_0$.  
We then write 
the factor $A$ in terms of a total (gas + dust) core mass,
$M$, the dust mass absorption coefficient $\kappa_0$
(evaluated at $\nu_0$), and the distance to the object, $d$
(which is here taken as 700~pc for all cores):
\begin{equation}
A = \frac{M \kappa_0}{R_{\rm gd} d^2}.
\label{eq:mass}
\end{equation}
Since $\kappa_0$ refers to a dust mass, the gas-to-dust mass ratio,
$R_{\rm gd}$, is required in the denominator to infer total masses.  We adopt
$\kappa_{0} = 16$\,cm$^2$\,g$^{-1}$, evaluated at
$\nu_0=c/250$\,\micron. 
While the actual value of $R_{\rm gd}$ may depend on the physical conditions of the local 
ISM or molecular cloud (e.g., \citealp{frisch2003}, \citealp{vuong2003}, 
and references therein) we adopt the value of $R_{\rm gd} \simeq 100 $ for simplicity.
Equation~(\ref{eq:sed}) is fit to all of the BLAST and archive photometry, from
1200 to 70\,\micron\ using $\chi^2$ optimization. Color correction of the BLAST
flux densities, before comparing them with other photometry, is described 
in \citet{chapin2008} and \citet{truch2008}, 
and the resulting BLAST color-corrected fluxes are listed in
Table~\ref{tab:src}. 
$\beta$ is not well constrained and we choose to fix it to $\beta=2.0$
(see \citealp{netterfield2009} for a discussion about the adopted values for $\kappa_{0}$ and $\beta$),
while only $A$ and $T$ are allowed to vary. 
%

%
 \begin{figure}
 \centering
 \includegraphics[width=3in]{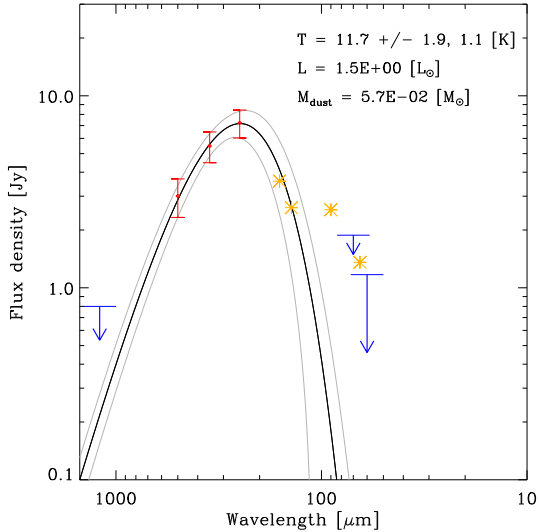}
 \includegraphics[width=3in]{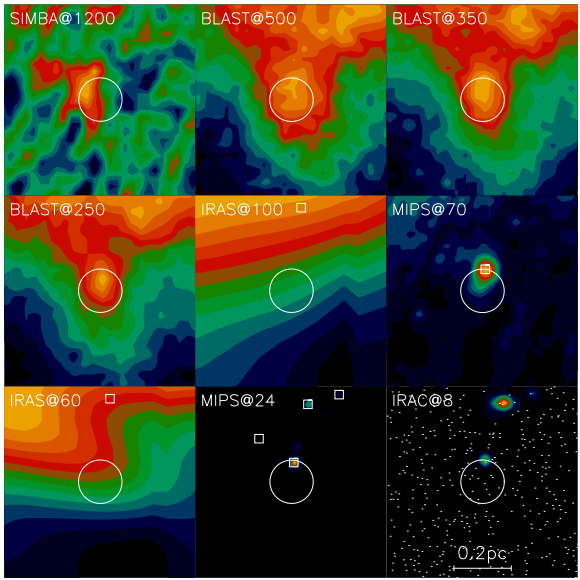}
 \caption{Same as Figure~\ref{fig:SEDex1} for one of the coldest sources
in Vela-D, BLAST J084542-432721.
   }
\label{fig:SEDex2}
\end{figure}

``Survival analysis'' is adopted to properly include the upper-limits in 
the calculation of $\chi^2$ (see the discussion in \citealp{chapin2008}),
and we choose to use all photometry at wavelengths shortward of 
70\,\micron\ as upper-limits. This is motivated by the following reasons:
 (i) emission at 24\,\micron\ and shorter is coming from volumes of
material quite different from those emitting at longer wavelengths,
which is typically associated with the coldest and less dense dust; 
(ii) at shorter wavelengths there can be significant contributions from
hotter dust grains, PAH bands and other spectral features;
and, (iii) in proto-stellar
regions multiple sources may be present that are simultaneously associated
with the emission in the FIR/submillimeter wavebands. In the latter case, these
multiple fluxes are all summed together and taken as upper-limits
in the SEDs.

Figure~\ref{fig:SEDex1} shows an example SED for the source BLAST J084848-433225, 
one of the brightest BLAST objects in the sample, with a best fit temperature to the 
mm-FIR data of 20.7\,K. 
Likewise, Figure~\ref{fig:SEDex2} shows the SED of BLAST J084542-432721, 
one of the coldest objects in Vela-D ($T=11.7$\,K). As seen in this example,
observations across the BLAST wavelength range clearly reveal a
turnover in the FIR SED for the coldest objects.
The thumbnails shown in the bottom panel of Figure~\ref{fig:SEDex2} clearly emphasize
the lack of counterparts in the shorter-wavelength bands for this object, though
it is located near a warmer object with clear MIPS24 and MIPS70 counterparts.

Particularly for the coldest objects, undetected in the FIR-MIR
wavebands, there is a strong degeneracy between the values of $\beta$ and $T$,
i.e. there can be a large spread in the values of these parameters that
can fit the SED equally well . This effect is less important for more evolved
sources, which usually have a FIR-MIR counterpart, thus better
constraining the SED shape and location of the FIR peak.
Fixing the value of $\beta=2.0$
has the advantage of reducing the inferred errors in $T$ \citep{chapin2008}
and makes the shape of the SED depend on temperature only,
but it clearly prevents the analysis of possible variations of $\beta$
as a function of other source parameters.
The physical parameters derived from the best-fit SED for each source are listed in
Table~\ref{tab:src}, whereas the flux densities used in the SED and the counterparts found
in each catalog are listed in Table~\ref{tab:flux}.

%
 \begin{figure*}
 \centering
 \includegraphics*[width=18cm]{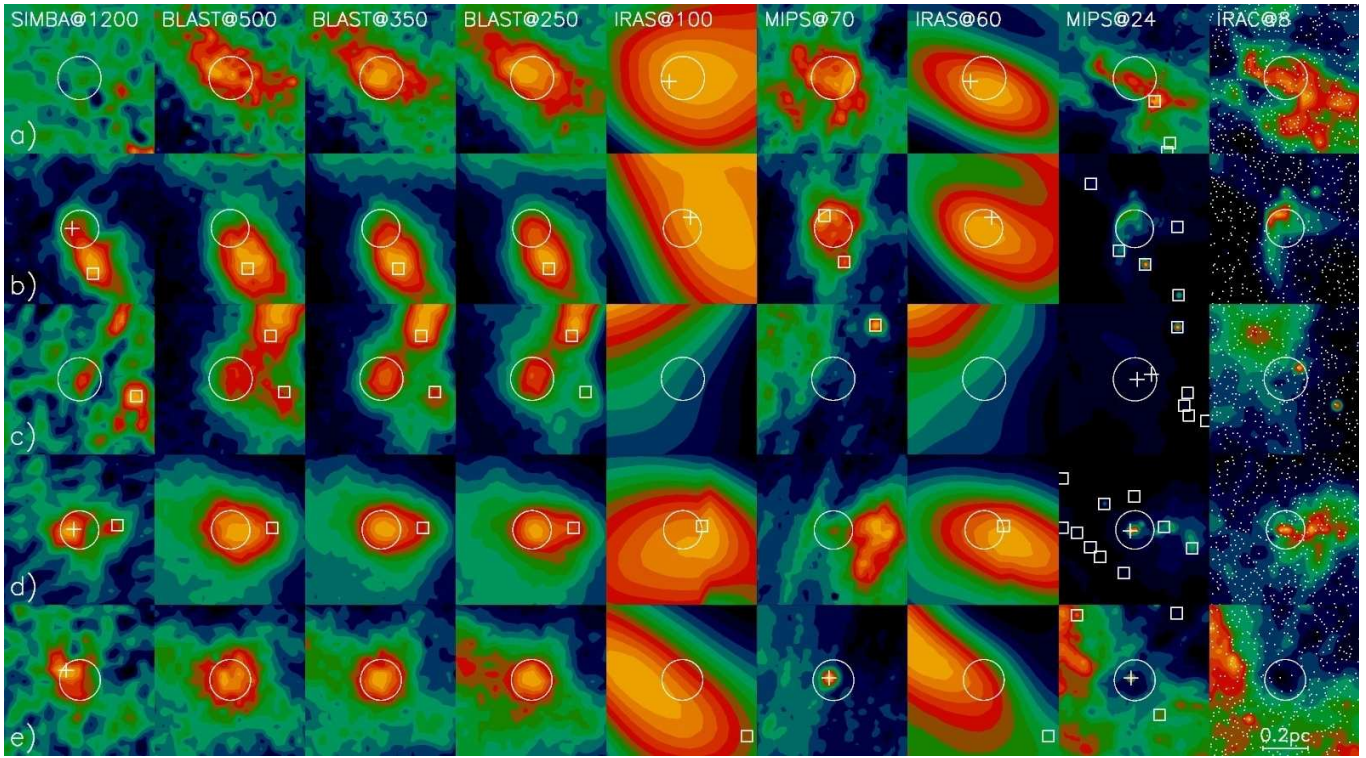}
%
 \caption{ Rows (a) and (b) show two examples of cores (BLAST J084805-435415 and BLAST J084902-433802,
respectively)
with no MIPS24 point source counterpart, but with MIPS70 compact emission.
Core BLAST J084805-435415 is indeed proto-stellar as it is found to be associated
with a molecular outflow (see text).
Rows (c) and (d) show two  cores (BLAST J084842-431735 and BLAST J085010-431704,
respectively) with
no compact MIPS70 emission but with  MIPS24 counterparts. Row (e) represents an
example of a cold BLAST core (BLAST J084928-440426) with both MIPS24 and MIPS70 counterparts.
Wavebands are ordered from left to right as follows:
SIMBA, BLAST500, BLAST350, BLAST250, IRAS100, MIPS70, IRAS60, MIPS24 and IRAC8.
Symbols have the same meaning as in Figure~\ref{fig:SEDex1}.
   }
\label{fig:SEDex3}
\end{figure*}

\section{COMPARING THE PHYSICAL PROPERTIES OF STARLESS AND PROTO-STELLAR CORES}
\label{sec:comparison}

\subsection{Separating starless and proto-stellar cores}
\label{sec:separating}

In this section we consider the properties of starless and proto-stellar cores.
The former will be defined as those cores without an internal source of luminosity
(i.e., a protostar), and may be further defined as ``pre-stellar'' when they
appear to be gravitationally bound. Proto-stellar cores, on the
other hand, have an embedded self-luminous source (which may also be a cluster
of objects) and possibly show other signs of SF, such as molecular outflows.

Starless and proto-stellar cores may be differentiated by identifying MIR
counterparts associated with the submillimeter cores.
Here we define as proto-stellar those cores which have  
a MIPS24 compact source, with flux greater than 3\,mJy (see \S\ref{sec:mips24irac}), 
within the search radius defined in \S\ref{sec:phot}.
For the BLAST cores outside the area covered by the MIPS24 map  (a total of 52 objects) we use
the association with compact sources from the {\it Midcourse Space Experiment} ({\it MSX}) catalog,
although the sensitivity limit is much higher than that of the MIPS24 catalog
(e.g., \citealp{egan1999}), and thus some of the starless cores outside the MIPS 
covered area might have a yet undected MIR counterpart (see \S\ref{sec:bestfittemp}, 
\S\ref{sec:bestfitother} and \S\ref{sec:bias} for a discussion of this point). 
We note that some of the point-like objects detected by MIPS24 may have 
a very low luminosity,  so we would expect that they neither
heat up the core overall nor increase its luminosity. Thus, the emission of these 
{\it early} proto-stellar cores 
in the BLAST wavebands will look much the same as the starless cores. 

As SF in the core proceeds, more and more luminosity is generated, mostly
from accretion but also from the quasi-static contraction of the interior and, later,
from nuclear fusion. This luminosity is re-processed by dust and hence these objects
will appear more luminous and warmer (and thus also with a larger luminosity-to-mass ratio, 
$L/M$). In terms of the observational properties, 
one should therefore expect a smooth transition from the pre- to the proto-stellar 
phase. Our criteria are not meant to
provide a sharp separation between these two classes of objects, but should provide guidelines
for understanding the physical conditions where the transition takes place.

\subsection{MIPS24 and IRAC counterparts}
\label{sec:mips24irac}

Analysing in more detail the association of MIPS24 point-sources with the BLAST cores,
we find that only two cores, BLAST~J084612-432337 and BLAST~J084745-432637,  
have MIPS24 counterparts with fluxes lower than 3\,mJy, i.e., the threshold selected 
by \citet{enoch2008} to eliminate most extragalactic interlopers. We thus classify these
two cores as starless.
By comparing the surface density of MIPS24 catalog sources in the area covered by the MIPS24 data with 
a circular area of radius equal to the search radius, we estimate
the probability of finding a chance association with a MIPS24 source, brighter than 3\,mJy,
within the search radius 
to be less than 10\%. 
Thus, although chance associations of BLAST cores 
with MIPS24 point sources seem unlikely, in \S\ref{sec:bias} we analyse how these might
potentially affect the overall distribution of physical parameters between starless and
proto-stellar objects.

%
%
%
\begin{figure*}
\centering
\includegraphics[width=8cm]{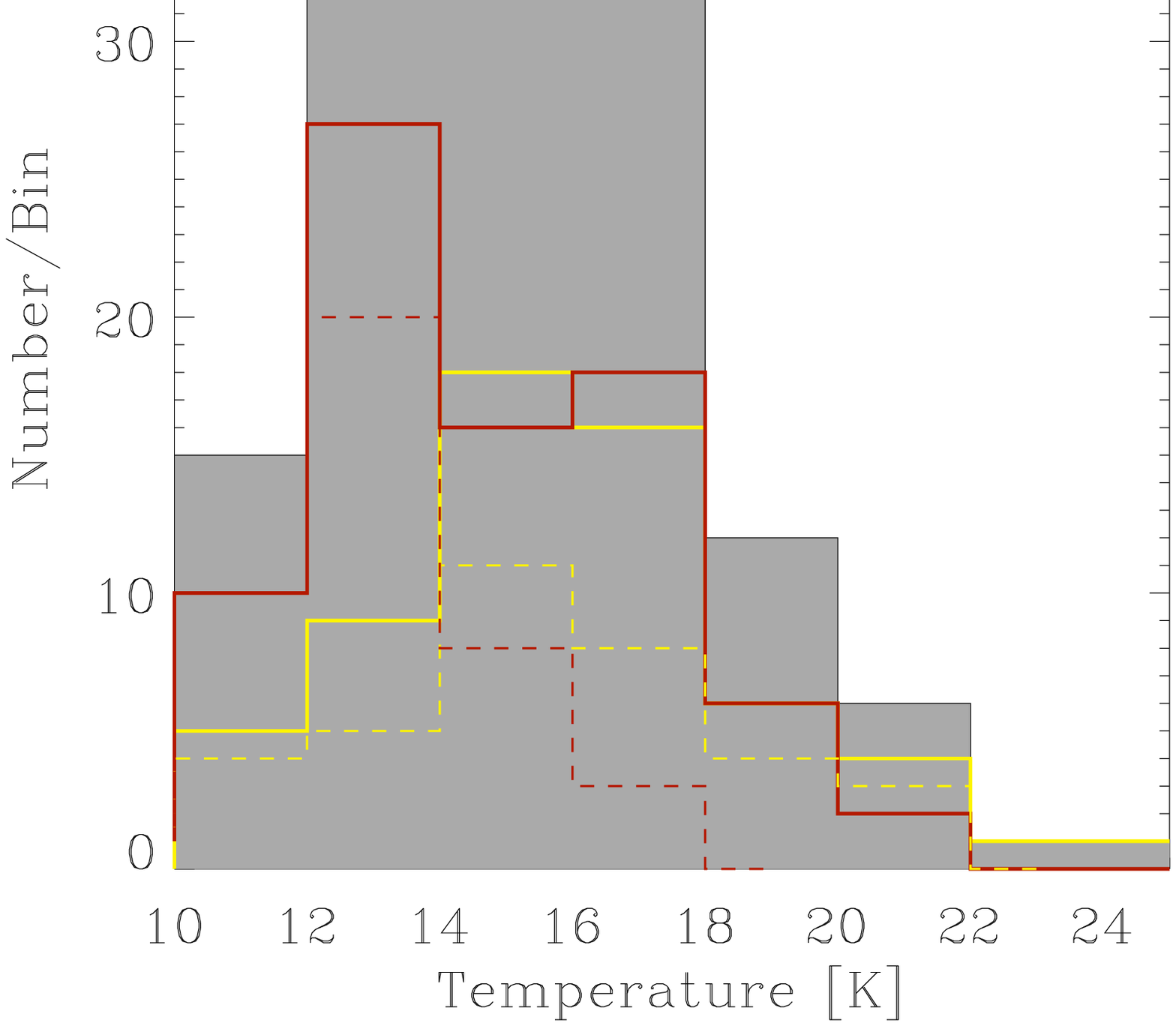}
\includegraphics[width=8cm]{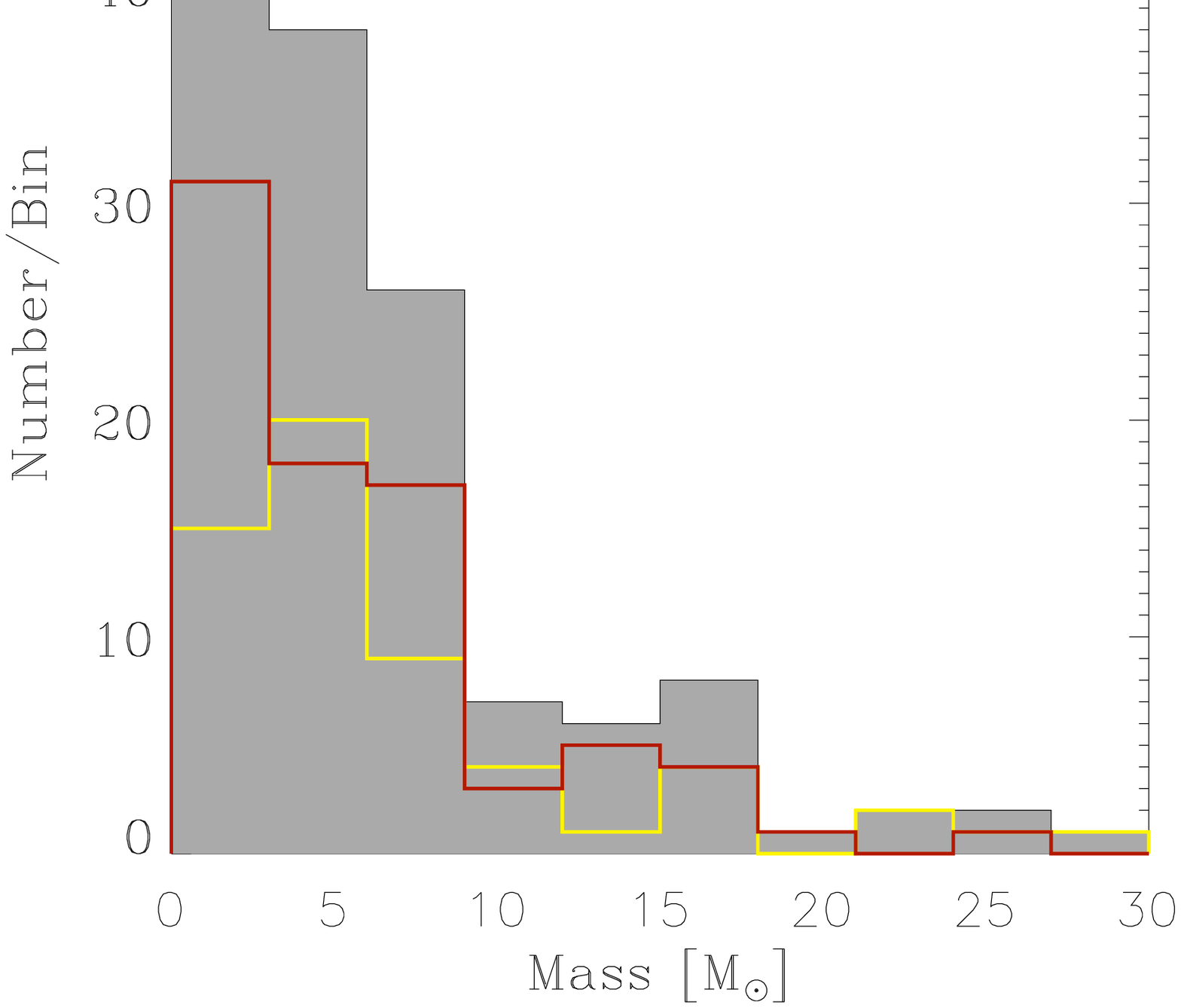}
\includegraphics[width=8cm]{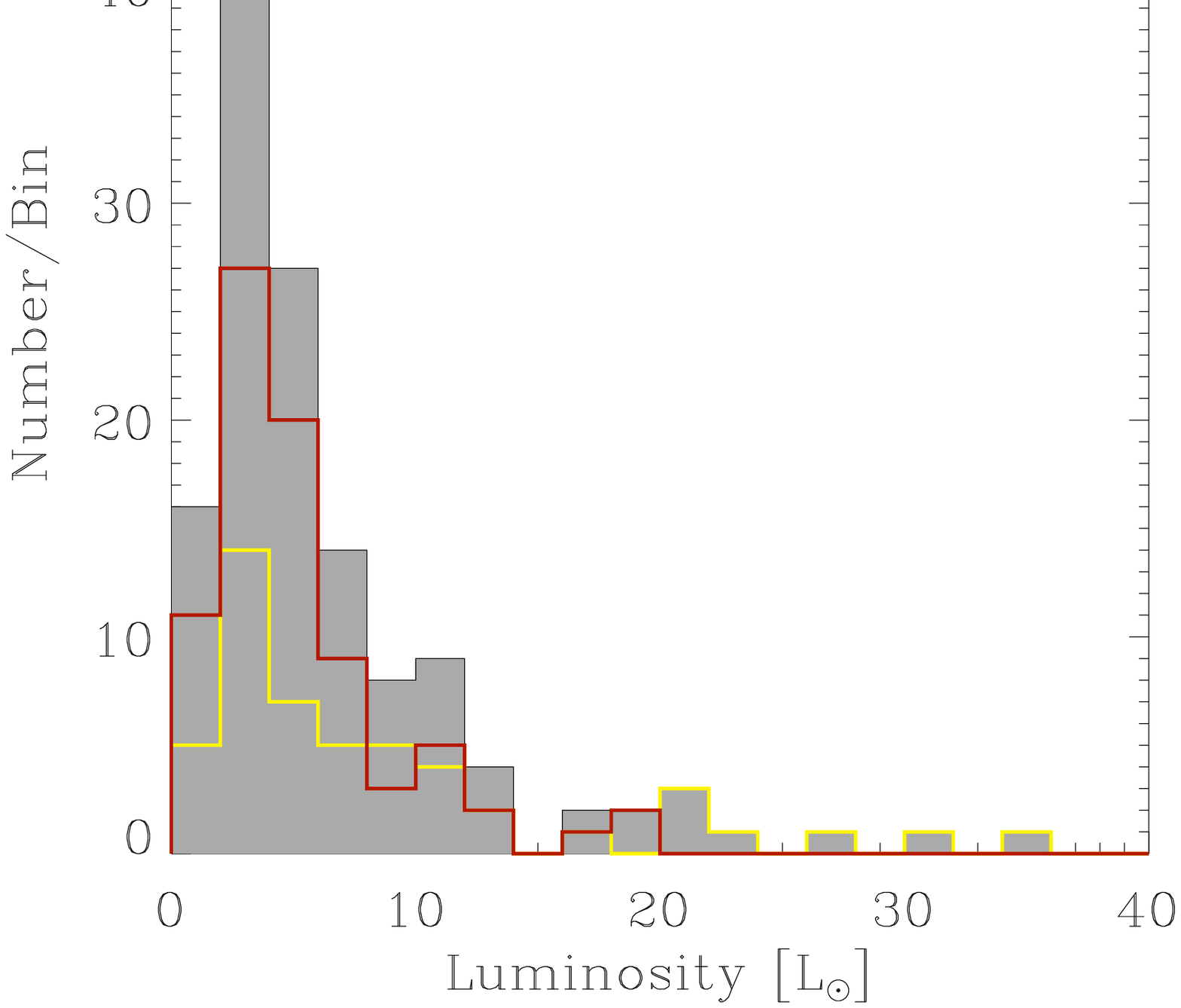}
\includegraphics[width=8cm]{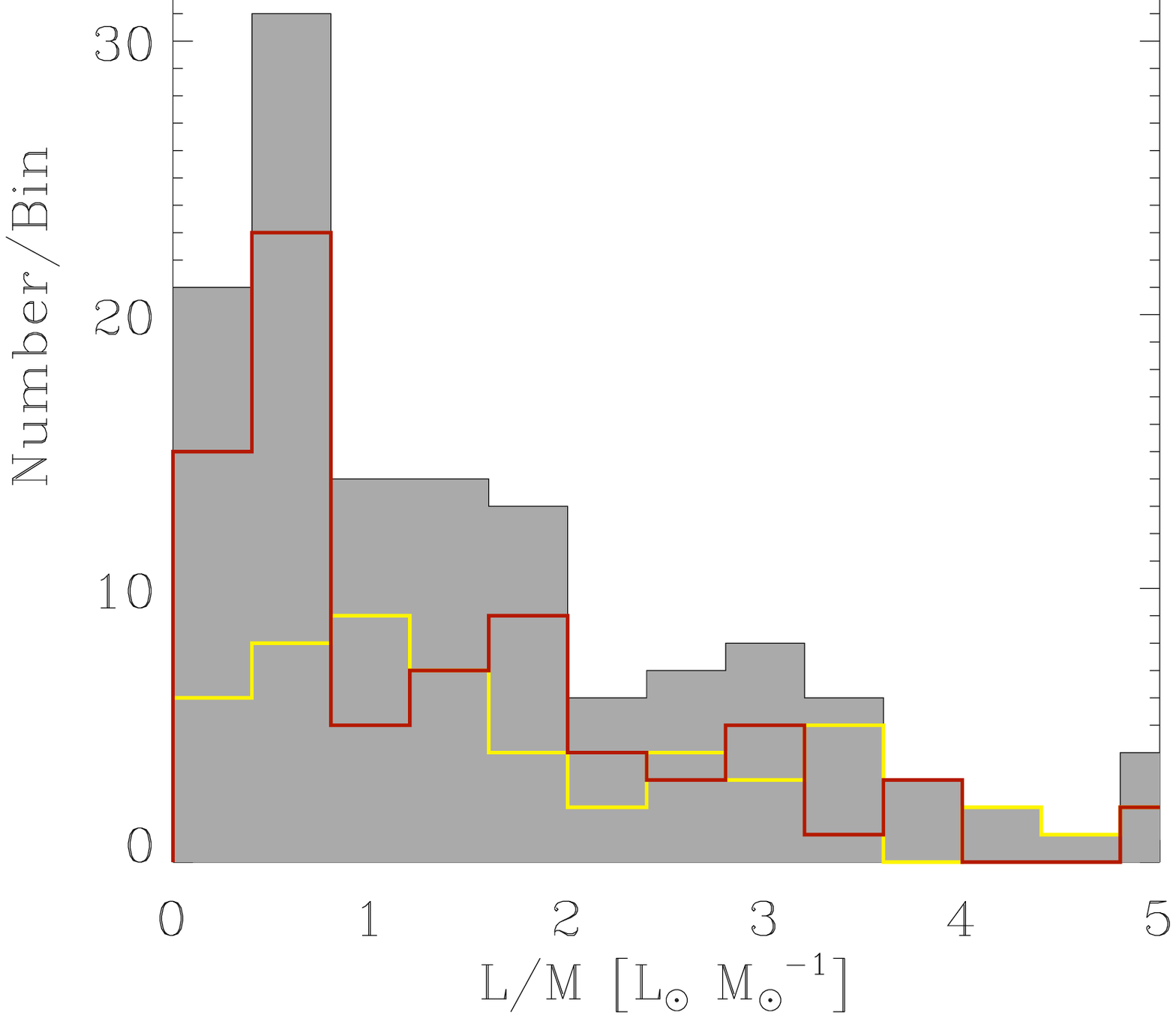}
\caption{
Overall distributions of physical parameters of the BLAST cores in the Vela-D region shown in
Figure~\ref{fig:blast250} and defined in \S\ref{sec:bestfittemp} and \S\ref{sec:bestfitother}
 (shaded histograms).
Proto-stellar (yellow solid lines) and starless cores (red solid lines) are defined in the text.
Shown are the distribution of the temperatures ({\it top-left panel}), which also includes
the distributions for cores with $M > 4.2\, {\rm M}_{\odot}$ (dashed lines; see Table~\ref{tab:prevsproto}),
masses ({\it top-right panel}),  luminosities ({\it bottom-left panel}) and
luminosity-to-mass ratios ({\it bottom-right panel}).
  }
\label{fig:mass_lum_histo}
\end{figure*}

\begin{deluxetable*}{llccccccc}
\tablewidth{0pt}
\small
\tablecaption{Median Values of Main Physical Parameters of Starless and Proto-stellar Cores
in Vela-D \label{tab:prevsproto}}
\tablehead{
\colhead{} &
\colhead{} &
\multicolumn{3}{c}{{\bf Starless}} &
\colhead{} &
\multicolumn{3}{c}{{\bf Proto-stellar}} \\
\colhead{} &
\colhead{} &
\colhead{All} &
\colhead{$M>4.2\, {\rm M}_\odot$} &
\colhead{$M>11\, {\rm M}_\odot$} &
\colhead{} &
\colhead{All} &
\colhead{$M>4.2\, {\rm M}_\odot$} &
\colhead{$M>11\, {\rm M}_\odot$}
}
\startdata
 $T$           & [K]                   & 14.6  & 13.1  & 12.9  &   & 15.7  & 15.5  & 17.2   \\
 $M$           & [${\rm M}_{\odot}$]   & 4.6   & 7.0   & 15.3  &   & 4.8   & 8.7   & 16.5   \\
 $L_{\rm FIR}$ & [${\rm L}_{\odot}$]   & 4.4   & 4.5   & 7.0   &   & 7.5   & 10.9  & 84.7   \\
 $L_{\rm FIR}/M$  & [$ {\rm L}_{\odot} {\rm M}_{\odot}^{-1}$]   & 1.0   & 0.5   & 0.5   &   & 1.6   & 1.5   & 2.7    \\
\enddata
\tablecomments{Median values of the best fits to the SEDs.
The two completeness limits, $M>4.2\, {\rm M}_\odot$ and $M>11\, {\rm M}_\odot$, correspond to sources
with $T>12$~K and $T>10$~K, respectively.
}
\end{deluxetable*}

We also find cases (a total of 26 objects) where a source is detected in the MIPS24 band, 
but {\it not} in the MIPS70 band. 
As an example, we show in Figure~\ref{fig:SEDex3}(c) source BLAST J084842-431735, 
where we find an associated point source from the MIPS24 catalog, but no point or even 
compact MIPS70 emission at the position of the BLAST core. In more than one third of other similar cases,
some compact weak or halo emission associated with the BLAST core is found in the MIPS70 map.
An example is shown in Figure~\ref{fig:SEDex3}(d): we can clearly see a weak compact source  
of emission in the MIPS70 map at the position of the BLAST core, though no point source is 
found in the catalog. In such cases, it is likely that a combination of intrinsically low 
70\,\micron\  emission and/or instrumental artifacts (e.g., stripes) in the MIPS70 maps may
mask the 70\,\micron\  counterpart to the MIPS24 source associated with the BLAST core.
 Geometrical projection effects may also be playing an important role at wavelengths $\la 100$\,\micron\ 
for sources with an embedded protostar (e.g., \citealp{whitney2004}). 

Identifying the NIR/MIR young proto-stellar counterparts associated to the BLAST cores 
is beyond the scope of this work. However, we have carried out a preliminary search of
IR candidates among the IRAC sources, selected using the following criteria: (i) spectral index
$\alpha_{\rm sp} = d \, {\rm log}(\lambda S_{\lambda}) / d \, {\rm log}(\lambda)$, estimated
using 2MASS, IRAC and MIPS24 data, with values $\alpha_{\rm sp} > 0.3$; (ii) 
the source must be detected in the IRAC 8\,\micron\ band; and (iii) when no MIPS24 flux
is available, the flux extrapolated from 2MASS and IRAC data must be consistent with 
an upper limit in the MIPS24 band. With these criteria, we find that 69\% of the starless
cores, as defined above,  within the area in Vela-D covered by IRAC
do {\it not} have an IRAC candidate. Likewise, we find that 73\% of the proto-stellar
cores do have an IRAC candidate. Other criteria can be found in the literature 
(see, e.g., \citealp{delu07}, \citealp{enoch2009}), but our quick search indicates that
the starless and proto-stellar populations have been fairly accurately characterized.

However, we can also use other signposts of early SF activity, such as proto-stellar jets and
molecular outflows.
In fact, we have identified 13 BLAST cores associated with the H$_2$ jets and knots observed by
\citet{gianni05}, \citet{gianni07} and \citet{delu07}, and all of them had been correctly
identified as proto-stellar, except the core BLAST J084822-433152, which has no MIPS24 (and no MIPS70)
counterpart and had thus been classified as starless. However, the association of this specific
core with an H$_2$ jet is not as robust as the others \citep{delu07}
and thus given its other properties we have decided to keep its classification as starless.
We have also identified the BLAST cores associated with the $^{12}$CO(1--0) molecular outflows
observed by \citet{elia2007} and found that only one core, BLAST J084805-435415, had been
previously incorrectly classified as starless.

As an additional category, we also 
find some BLAST cores that are detected in MIPS24, MIPS70 (and in the IRAC bands) that
are characterized by quite low temperatures. 
An example is source BLAST J084928-440426, 
shown in Figure~\ref{fig:SEDex3}(e), which is clearly associated with bright MIPS24 and 70
point sources. However its temperature, as determined from the best fit to the SED, is only 13.1~K. 
Such a cold source is not expected to emit
a detectable flux in the MIPS70 band and we are therefore left with two possible scenarios:
(i) either the 70\,\micron\  flux is mostly emitted by 
the warmer central source associated with
the BLAST core, and thus the BLAST fluxes represent the contribution from a much colder envelope;
or, (ii) the BLAST fluxes and the MIR emission are associated with two (or more) separate
objects, a cold, starless core detected by BLAST, and a nearby proto-stellar core detected
by MIPS with negligible emission in the submillimeter. Because of our inability to discriminate
between these two scenarios, and for consistency with the criteria given above, we consider
these cores as proto-stellar.

%
%
\begin{figure}
\centering
\includegraphics[width=7cm]{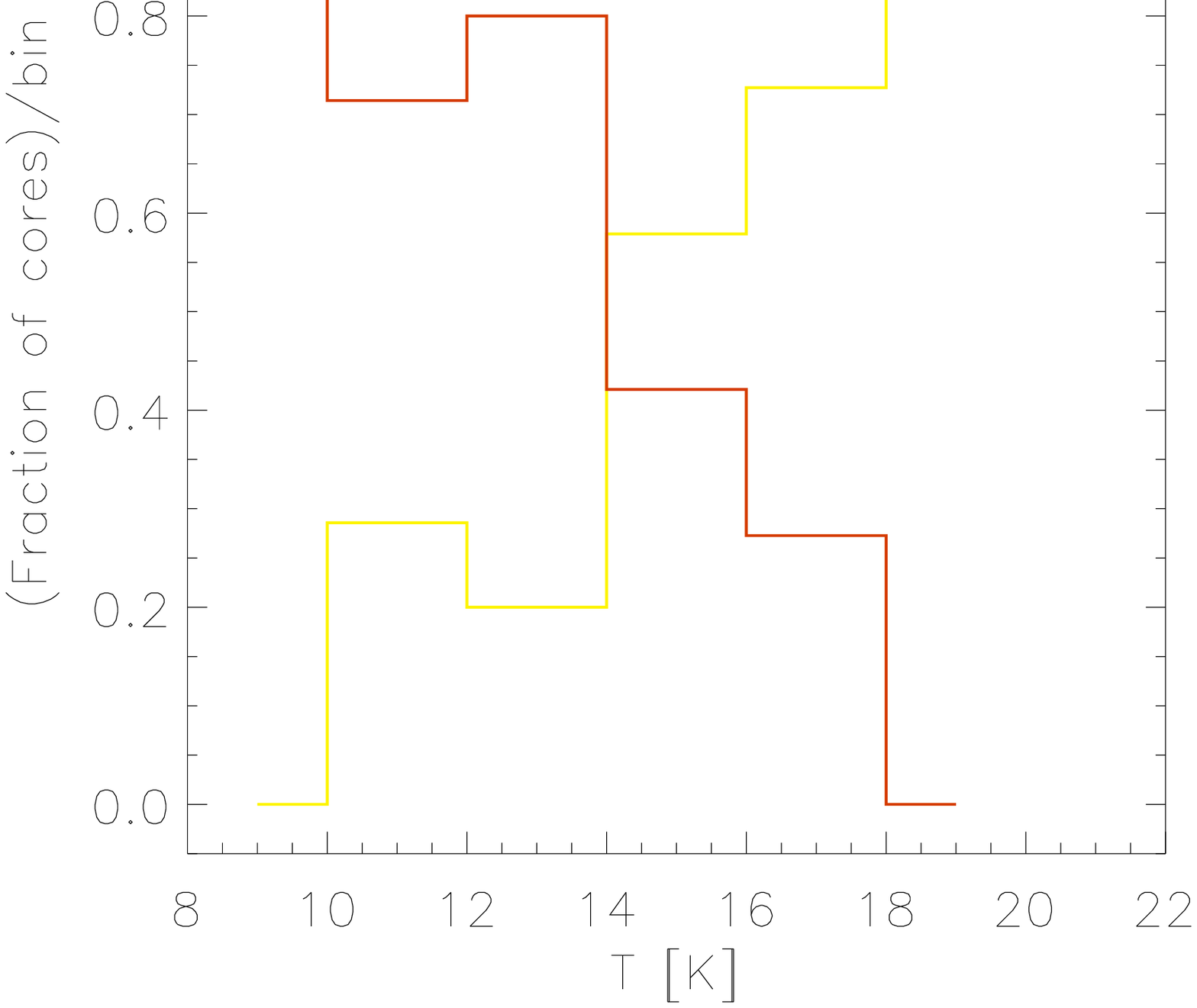}
\includegraphics[width=7cm]{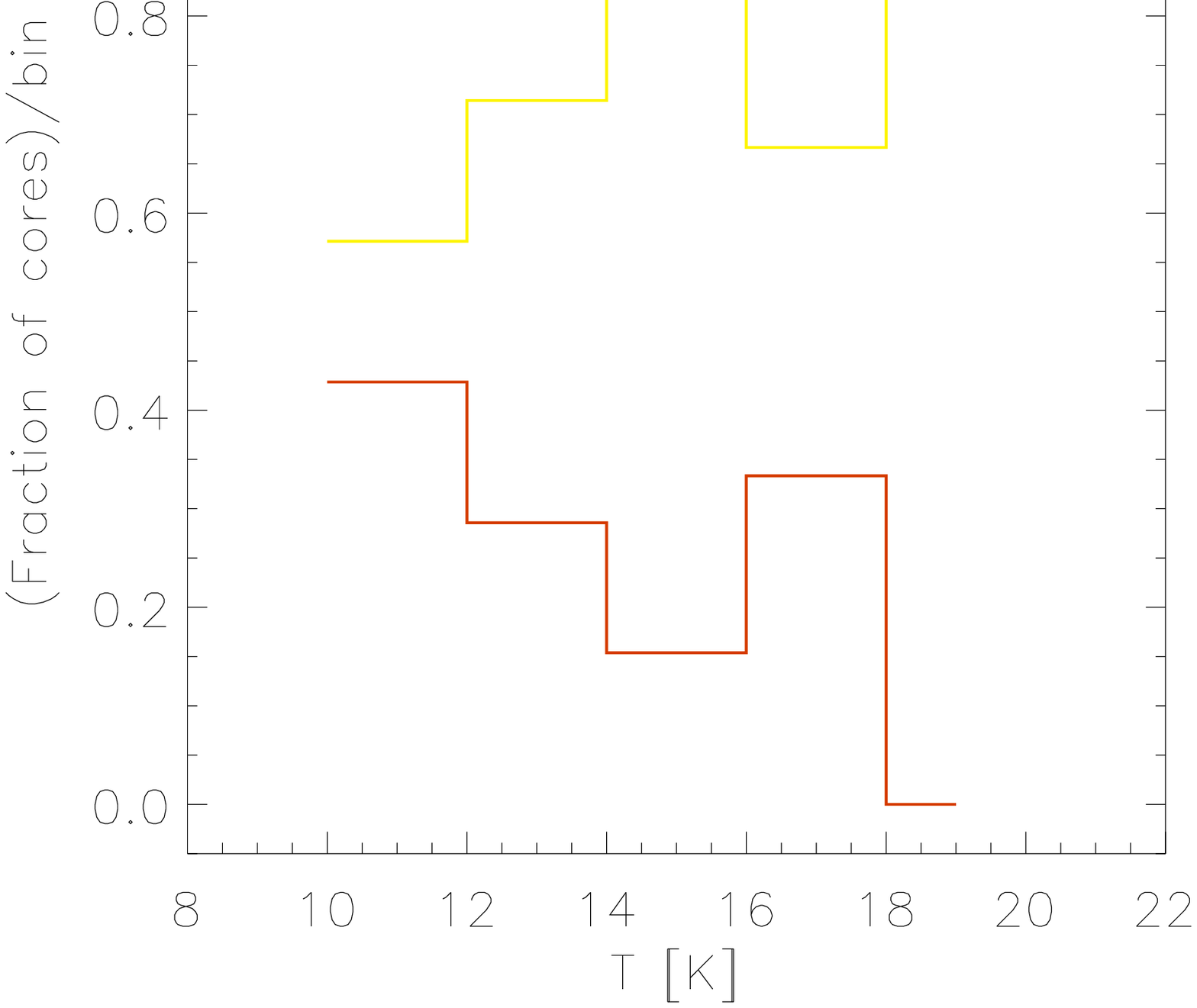}
\caption{
{\it Top.} Fraction of starless (red line) and proto-stellar (yellow line) cores vs.
temperature, in the whole Vela-D region and only for cores with $M>4.2\, {\rm M}_\odot$.
{\it Bottom.} Same as above in the region covered by the MIPS24 map  (see discussion
in \S\ref{sec:bestfittemp}  and \S\ref{sec:bias}).
  }
\label{fig:corefraction}
\end{figure}

Finally, we note that in objects such as core BLAST J084928-440426 
(of the previous example) although associated with point sources in the IRAC and MIPS
wavebands, the peak of the submillimeter emission corresponds to a minimum in
the MIPS24 and IRAC8 maps, as shown in Figure~\ref{fig:SEDex3}(e). 
This and other similar cores  
stand out as dark shadows against the diffuse 8 and 24\,\micron\  emission.

\subsection{Distribution of physical parameters}
\label{sec:masslum}

\subsubsection{Best fit values: temperature}
\label{sec:bestfittemp}

Temperature distributions for starless and proto-stellar cores in Vela-D
are shown in Figure~\ref{fig:mass_lum_histo}. The histograms refer to the values
of the physical parameters as obtained from the best fit to the SED of each source
found in the area defined in \S\ref{sec:blastobs} and Figure~\ref{fig:blast250}.
The effects, on the overall distribution, of the errors on the parameters 
determined for each individual source will be discussed in \S\ref{sec:MCsim}.
It must also be noted that only 89 objects are located within the common
area covered by the MIPS24 and MIPS70  maps (see Figure~\ref{fig:map_coverage}).

In Figure~\ref{fig:mass_lum_histo} we note that
the temperature distribution has a wide peak 
at $T \sim 15$\,K, with a sharp cutoff at the low end around $11-12\,$K and also at the high end,
at about $18-20$\,K. In fact,  less than 7\% of the sources are warmer than 20\,K, and they are
mostly proto-stellar objects.
The median value for the whole sample is 15.4\,K,
and we note that proto-stellar cores are slightly warmer (median temperature 15.7\,K)
than starless cores (median temperature 14.6\,K).
However, if we restrict ourselves to mass ranges for which we are progressively more complete, 
then the temperature difference between starless and proto-stellar cores appears to 
become increasingly more significant, as shown in Table~\ref{tab:prevsproto} and by the dashed lines 
in the top-left panel of Figure~\ref{fig:mass_lum_histo}. Clearly,
in this case the sample size also becomes smaller.
The higher value of the overall median temperature,
as compared for example to the Pipe cores \citep{rathborne2008} or the cores in Vela-C, 
suggests that the cores in Vela-D are indeed in a later stage
of evolution. The range of values found for the spectral index (\S\ref{sec:mips24irac}) 
is an evidence for different evolutionary stages within the proto-stellar population.
However, we have not attempted to separate the starless and proto-stellar populations
into further sub-groups.

One question of interest is whether temperature is a good discriminator of starless and
proto-stellar cores, in a statistical sense. The overlap between the temperature distributions
of starless and proto-stellar cores  in Figure~\ref{fig:mass_lum_histo}, 
besides being affected by completeness effects (see Table~\ref{tab:prevsproto}), is also
likely to be a consequence of the smooth transition from the pre- to the
proto-stellar phase, as discussed in \S\ref{sec:separating}.
This can also be seen in the top panel of Figure~\ref{fig:corefraction}, where we show the 
fraction (out of the total number of cores in each bin) 
of starless and proto-stellar cores, as a function of core temperature, in the whole Vela-D region. 
The plot clearly shows that the fractions
of starless and proto-stellar cores have opposite trends with temperature. 
This trend is also visible, though to a lesser extent, in the smaller region covered by the MIPS24 map
(bottom panel of Figure~\ref{fig:corefraction}), where we are able to better discriminate starless 
from proto-stellar cores (\S\ref{sec:separating}).  If sensitive MIR observations were also available
outside the MIPS24 map one would expect the top panel of Figure~\ref{fig:corefraction} to become 
more like the histogram in the bottom panel. Possible bias effects caused by the use of MIPS24 and MSX data in
different regions of Vela-D are later discussed in \S\ref{sec:bias}.

%
%
\begin{figure}
\centering
\includegraphics[width=8cm]{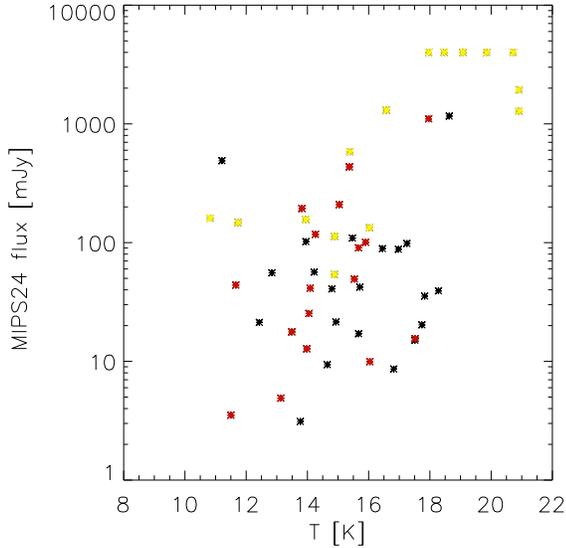}
\caption{
MIPS24 flux of the proto-stellar cores vs. core temperature.
The red and yellow symbols represent the complete sample of cores, with mass
in the ranges $4.2\, < M < 9 \, {\rm M}_\odot$,
and $ M > 9 \, {\rm M}_\odot$, respectively. The black symbols represent cores
with masses $< 4.2\, {\rm M}_\odot$.
Sources with flux $\ge 4000 \, $mJy are saturated (\citealp{gianni07}, \citealp{stra2009}).
  }
\label{fig:f24vstemp}
\end{figure}

The distribution of temperature among the Vela-D sources can also be used to search
for evidence of evolution within the proto-stellar cores found in the MIPS24 map. 
Thus, in Figure~\ref{fig:f24vstemp} we plot the MIPS24 flux as a 
function of core temperature, and we also separate cores in two different mass ranges, 
to be able to partially disentangle evolutionary effects from core mass-related effects. 

Figure~\ref{fig:f24vstemp} then shows two distinct features: (i) first, there is a trend for
the MIPS24 flux to increase with temperature; and, (ii) this trend is visible also
{\it within} the two selected mass ranges, though these two subsets appear to populate
different regions of the plot. The sources with higher masses, in fact, appear to  
also have somewhat higher MIPS24 fluxes and temperatures. 
We tentatively explain the main trend of flux vs. temperature as an evidence of
different {\it evolutionary} phases within the population of proto-stellar cores, while 
the partial segregation of the two subsets described above appears to be associated
with the mass of the core.  This result is consistent with the trend in the 
$L_{\rm bol} - T_{\rm bol}$ diagram observed by \citet{enoch2009} and modeled 
by \citet{young2005} (see also \citealp{whitney2003}). The increase in $L_{\rm bol}$ is accompanied
by a general increase of the flux density at MIR wavelengths.

\subsubsection{Best fit values: mass and luminosity}
\label{sec:bestfitother}

The overall mass distribution in Figure~\ref{fig:mass_lum_histo} shows,
as expected,  an almost monotonic decrease from lower to higher masses.
It also shows very few cores with $M > 10\, {\rm M}_{\odot}$. In fact,
the median value for the whole sample is $4.7 {\rm M}_{\odot}$, whereas from
Table~\ref{tab:prevsproto} we note that
the median masses of starless and proto-stellar cores are not significantly different,
suggesting that the envelope mass is not severely affected by the transition between 
the two phases.

%
%
\begin{figure}
\centering
\includegraphics[width=7cm]{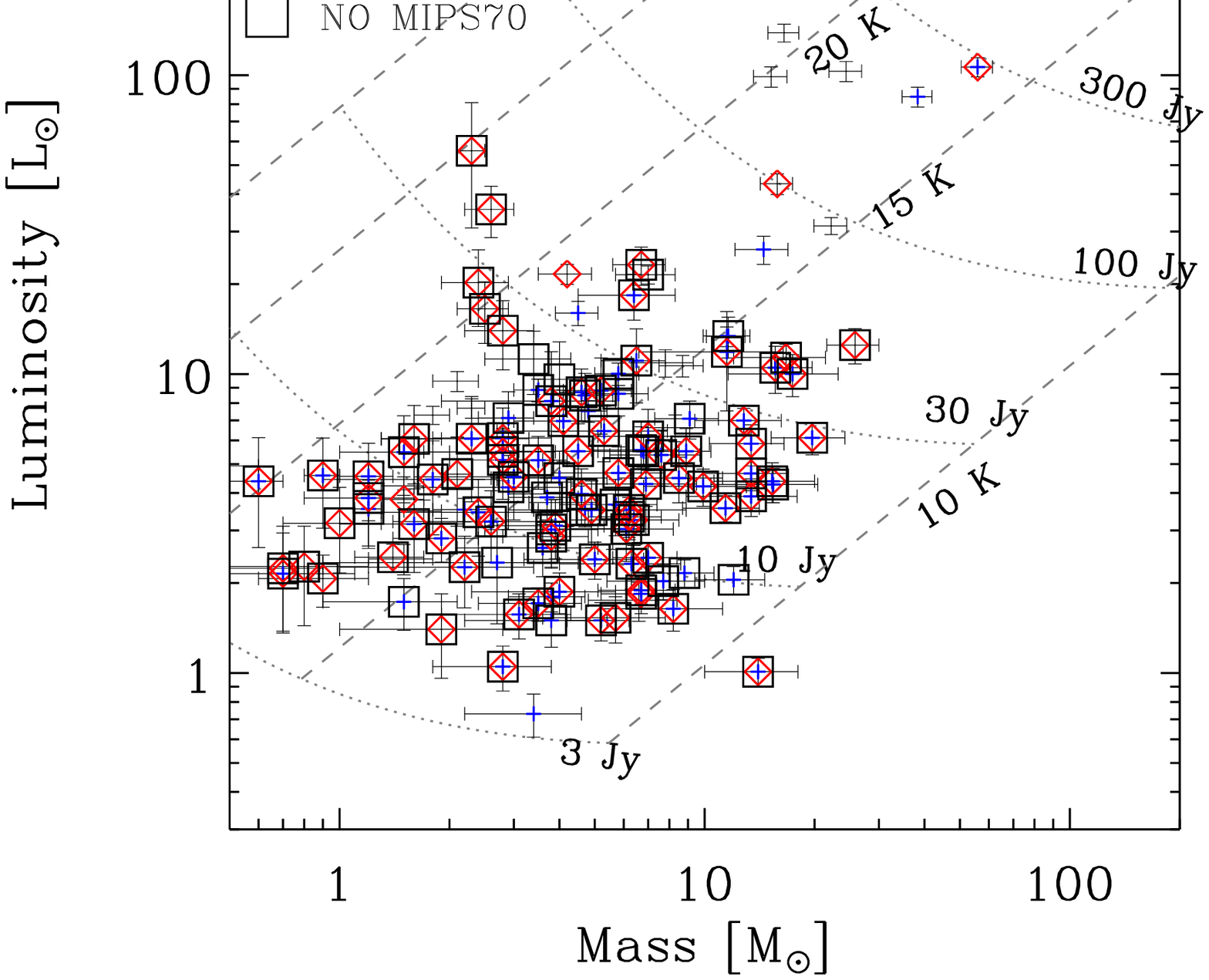}
\includegraphics[width=7cm]{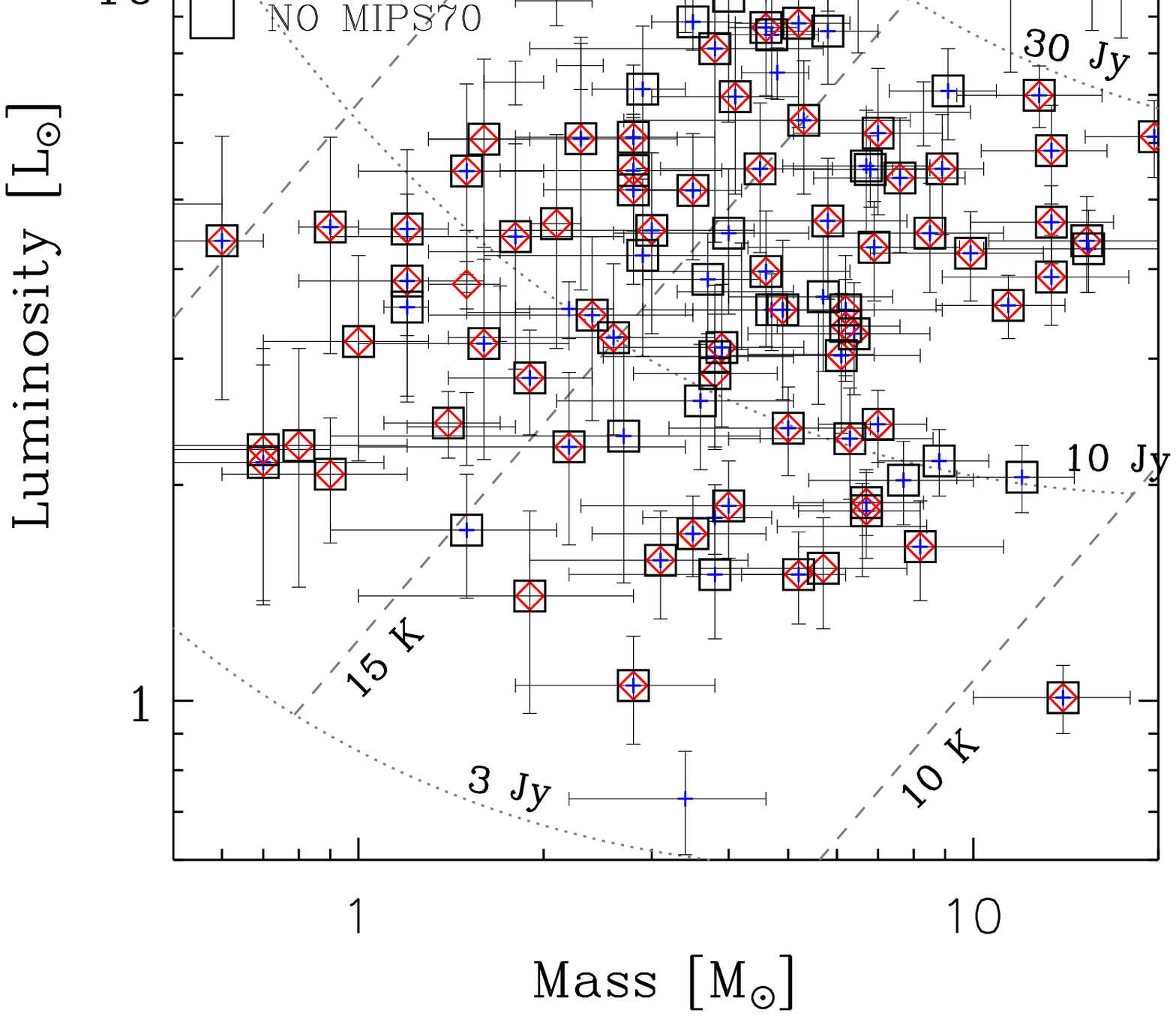}
\caption{
{\it Top.} Luminosity versus mass for the BLAST sources in Vela-D.
  The errors shown in mass and luminosity are
  estimated from the range of temperatures consistent with the BLAST
and archival photometric data under the assumption of the simple SED model of
  Equation~(\ref{eq:sed}).  We do not include additional sources of error
  involving the distance to Vela-D and the adopted values of
  $\beta$, $\kappa_0$, and gas-to-dust mass ratio.  The dashed lines
are loci at constant $T=10$, 15, 20, 25 and 30\,K,
corresponding (assuming a modified blackbody SED with $\beta=2.0$)
 to constant $L_{\mathrm{FIR}}/M$
values of 0.07, 0.65, 3.2, 10.9 and 29.6\,${\rm L}_\odot\,{\rm M}_\odot^{-1}$
  Roughly orthogonal to these are loci (dotted lines) of constant 250\,\micron\
  flux density, ranging from 3 to 1000\,Jy, using the same model.
{\it Bottom.} Zoomed plot at low masses and luminosities.
  }
\label{fig:lumvsmass}
\end{figure}

%
\begin{deluxetable*}{llccccc}
\tablewidth{0pt}
\small
\tablecaption{Average Values from Monte Carlo Simulations
for All Cores in Vela-D. \label{tab:prevsprotoALL}}
\tablehead{
\colhead{} &
\colhead{} &
\multicolumn{2}{c}{{\bf Starless}} &
\colhead{} &
\multicolumn{2}{c}{{\bf Proto-stellar}} \\
\colhead{} &
\colhead{} &
\colhead{mm-MIR} &
\colhead{BLAST} &
\colhead{} &
\colhead{mm-MIR} &
\colhead{BLAST}
}
\startdata
 $T$    & [K]                  & $13.1\pm 0.3$  & $13.1\pm 0.3$  &   & $15.6\pm 0.2$   & $15.7\pm 0.5$ \\
 $M$    & [${\rm M}_{\odot}$]  & $8.5\pm 0.8$   & $8.7\pm 0.9$   &   & $9.1\pm 0.8$    & $9.7\pm 1.1$  \\
 $L_{\rm FIR}$    & [${\rm L}_{\odot}$]        & $4.8\pm 0.3$   & $4.9\pm 0.7$   &   & $11.6\pm 0.8$    & $
13.1\pm 2.4$  \\
 $L_{\rm FIR}/M$  & [$ {\rm L}_{\odot} {\rm M}_{\odot}^{-1}$]   & $0.5\pm 0.1$   & $0.5\pm 0.1$   &   & $1
.5\pm 0.2$  & $1.4\pm 0.3$    \\
\enddata
\tablecomments{
The table lists the mean values obtained by averaging the medians of each Monte carlo simulation, 
for all cores in the whole Vela-D region. The columns labelled ``mm-MIR''
list the average values obtained using the SIMBA, BLAST and MIPS70 data in estimating the best-fit SED
of each individual source.  The columns
labelled ``BLAST'' list the values obtained from BLAST photometry only.
Only cores with $M>4.2\, {\rm M}_\odot$ (see Table~\ref{tab:prevsproto}) have
been selected.
}
\end{deluxetable*}

In Figure~\ref{fig:mass_lum_histo} we also show the FIR luminosity distribution
of the cores in Vela-D.  FIR luminosities, $L_{\rm FIR}$, are the integrated
luminosities from the modified blackbody fit. 
Though not shown (to make the histogram easier to read), 
we find 11 sources ($\simeq 8$\% of the total, all proto-stellar
but one, BLAST J084539-435133) with luminosities higher than $40 \, {\rm L}_{\odot}$.
The luminosity histogram shows a strong 
peak at $L_{\rm FIR} \sim 4 \,  {\rm L}_{\odot}$ 
and a rapidly decreasing trend at higher luminosities.
The median value for the whole sample is $4.6\,  {\rm L}_{\odot}$ 
and contrary to the mass distribution 
we  find a significant difference between the median luminosity of starless
and proto-stellar cores (see Table~\ref{tab:prevsproto}).

Although the luminosity-to-mass ratio, $L_{\rm FIR}/M$, is essentially equivalent to 
temperature (see later Figure~\ref{fig:lumvsmass}), it is useful to 
show it separately in Figure~\ref{fig:mass_lum_histo}
and in Table~\ref{tab:prevsproto}, as well as in the rest of the paper.  Like temperature, this ratio 
is also expected to increase with time as more and more gas is converted into stars
during the SF process and the embedded protostar, or proto-cluster,
becomes more luminous.  
For the simple SED model adopted, we find a range 
$L_{\rm FIR}/M \sim 0.2 - 40 \, {\rm L}_{\odot} {\rm M}_{\odot}^{-1}$, with $\simeq 10$\%
and $\simeq 3$\%
of the sources having $L_{\rm FIR}/M > 5\,  {\rm L}_{\odot} {\rm M}_{\odot}^{-1}$
and $L_{\rm FIR}/M > 10\,  {\rm L}_{\odot} {\rm M}_{\odot}^{-1}$, respectively.
The median value for the whole sample is $1.4\,  {\rm L}_{\odot} {\rm M}_{\odot}^{-1}$. 
Like temperature, the higher value of the luminosity-to-mass ratio 
found in proto-stellar cores does suggest that they are indeed at a later  
stage of evolution.

The temperatures and masses of the cores are also shown in Figure~\ref{fig:blast250}, for
both starless and proto-stellar cores, where we note some interesting features.
First, one can see the higher concentration of proto-stellar cores toward the central
filaments. While this might partly be a consequence of being unable to identify all
proto-stellar cores outside the MIPS24 area, we think that the presence here of many
IRS sources (\S\ref{sec:morph}) would indeed suggest that this region of Vela-D 
is in a later stage of evolution.  
We also tentatively note that the most massive cores are rarely found in isolation
(such as several of the IRS sources);
instead, they are often associated with small clusters of objects. On the other hand,
the most isolated cores tend to be low-mass objects. These characteristics, though 
no quantitative measurement is involved, seem to be in agreement with more systematic
observations (e.g., \citealp{faustini2009}) and numerical simulations (e.g., \citealp{smith2009}).
Another interesting feature
is the filament at $l\sim 262\fdg80$ and $b\sim +0\fdg3$ which is exclusively
composed of massive and cold starless cores.

Finally, in Figure~\ref{fig:lumvsmass} we show a plot of the luminosity as a function of mass. 
The plot shows a high concentration of points at temperatures $\la 15$~K,
or $L_{\mathrm{FIR}}/M \la 0.65\,  {\rm L}_\odot\,{\rm M}_\odot^{-1}$, which
clearly reproduces the results already shown in Figure~\ref{fig:mass_lum_histo}. 
Separated from the main concentration of sources, we note in Figure~\ref{fig:lumvsmass} 
a small group of 9 bright objects 
with flux density (at 250\,\micron)  $> 100$~Jy and $L_{\mathrm{FIR}} > 80 \, {\rm L}_\odot$. 
In fact, five of these correspond to known IRS sources \citep{Lis92}, 
IRS17 to IRS21, 
and all of these objects belong to small clusters of sources (\S\ref{sec:morph}). 
%

\subsubsection{Monte Carlo Simulations}
\label{sec:MCsim}

In this section we want to discuss the effects that the errors on the physical parameters
of each source, as determined using the procedure described in \S\ref{sec:SEDfit}, 
have on the overall statistical distribution. To do this, we have used Monte Carlo
simulations in which the values of temperature, mass, luminosity and luminosity-to-mass
ratio of each source are varied within the uncertanties listed in Table~\ref{tab:src}.
After each simulation, the histograms with the distribution of the four physical parameters
are evaluated and the median values are determined. Also, the values of each bin in the histograms
are stored for further computation. At the end of 500 such simulations, the mean and
standard deviation of the median values are determined and listed in Tables~\ref{tab:prevsprotoALL}
and \ref{tab:prevsprotoSM}.  

In order to analyze possible systematic effects and biases, during each simulation 
the random values of temperature, mass, luminosity and luminosity-to-mass ratio were first
determined based on the best fit results and uncertainties obtained using 
SIMBA, BLAST and MIPS70 photometry (\S\ref{sec:sed} and Table~\ref{tab:src}). 
Then, we repeated the procedure, but using the values obtained from BLAST-only photometry.
Furthermore, the Monte Carlo simulations have been performed for two different sets of data:
first, we used the complete sample of 141 Vela-D sources (see Table~\ref{tab:prevsprotoALL}). 
Then, we repeated the simulations only for those BLAST cores falling in both the SIMBA and MIPS70 
maps. In both cases we selected only those cores with $M>4.2\, {\rm M}_\odot$ 
(see Table~\ref{tab:prevsprotoSM}; see also note in Table~\ref{tab:prevsproto}), thus 
the values in these tables should be compared with columns 4 and 7 in Table~\ref{tab:prevsproto}.
In particular, Table~\ref{tab:prevsprotoSM} allows to evaluate the effects of our inability 
to correctly identify all proto-stellar cores outside the MIPS24 area (which overlaps well with 
the MIPS70 area, see Figure~\ref{fig:map_coverage}).
%

%
\begin{deluxetable*}{llccccc}
\tablewidth{0pt}
\small
\tablecaption{Average Values from Monte Carlo Simulations
for Cores in the SIMBA+MIPS70 Area. \label{tab:prevsprotoSM}}
\tablehead{
\colhead{} &
\colhead{} &
\multicolumn{2}{c}{{\bf Starless}} &
\colhead{} &
\multicolumn{2}{c}{{\bf Proto-stellar}} \\
\colhead{} &
\colhead{} &
\colhead{mm-MIR} &
\colhead{BLAST} &
\colhead{} &
\colhead{mm-MIR} &
\colhead{BLAST}
}
\startdata
 $T$              & [K]                        & $14.1\pm 0.9$  & $13.1\pm 1.0$  &   & $15.3\pm 0.2$   & $
15.1\pm 0.6$ \\
 $M$              & [${\rm M}_{\odot}$]        & $7.4\pm 1.2$   & $7.8\pm 1.6$   &   & $9.0\pm 0.8$    & $
9.7\pm 1.2$  \\
 $L_{\rm FIR}$    & [${\rm L}_{\odot}$]        & $5.0\pm 0.9$   & $4.1\pm 1.2$   &   & $9.9\pm 0.7$    & $
10.1\pm 1.7$  \\
 $L_{\rm FIR}/M$  & [$ {\rm L}_{\odot} {\rm M}_{\odot}^{-1}$]   & $0.8\pm 0.3$   & $0.5\pm 0.2$   &   & $1
.3\pm 0.1$  & $1.1\pm 0.2$
  \\
\enddata
\tablecomments{ Same as Table~\ref{tab:prevsprotoALL} for cores in the region of Vela-D covered by both
the SIMBA and MIPS70 maps. In addition, like Table~\ref{tab:prevsprotoALL}
only cores with $M>4.2\, {\rm M}_\odot$ have been selected.
}
\end{deluxetable*}

The results listed in Tables~\ref{tab:prevsprotoALL} and \ref{tab:prevsprotoSM} can be summarized
as follows: (i) in general, the uncertanties associated with the physical parameters obtained from 
BLAST-only photometry are higher, which is a direct consequence of the more accurate determination
of the physical parameters achieved through mm to MIR photometry (when available); (ii) the mean values
of the physical parameters obtained from BLAST-only and mm-MIR photometry are consistent,
within the errors, for both starless and proto-stellar cores, and separately in the two selected samples;
(iii) some differences may be seen between corresponding mean values of the whole sample and those
of the smaller SIMBA/MIPS70 sample, which reflect several effects such as:
completeness, correct identification of starless and proto-stellar cores outside the MIPS24 
coverage (\S\ref{sec:bias}), different evolutionary stage of the densest, central region in Vela-D
(\S\ref{sec:bestfitother}); (iv) despite these effects, the main features of starless and
proto-stellar cores discussed in \S\ref{sec:bestfittemp} and \S\ref{sec:bestfitother} 
are still observable in Tables~\ref{tab:prevsprotoALL} and \ref{tab:prevsprotoSM}.

\subsection{Potential effects biasing the distribution of physical parameters}
\label{sec:bias}

In \S\ref{sec:mips24irac} we mentioned that as many as 10\% of the MIPS24
counterparts to the BLAST cores could be the result of chance associations.
To estimate their potential impact on the distribution of physical parameters
between starless and proto-stellar cores we used the following procedure.
First, we randomly select 10\% of the proto-stellar cores and assume that they
are false associations, and thus transfer them into the sample of starless cores.
Then, we estimate the median of all physical parameters using these re-defined
samples of starless and proto-stellar cores. We repeat the previous two steps 500 times
and finally we estimate the {\it mean} of all the  median values previously determined.
We thus obtain for starless and proto-stellar cores, respectively:
$\langle T \rangle = 14.5\pm0.2$~K and $\langle T \rangle = 15.8\pm0.1$~K;
$\langle M \rangle = 4.7\pm0.2 \, {\rm M}_{\odot}$ and $\langle M \rangle = 4.8\pm0.3 \, {\rm M}_{\odot}$;
$\langle L_{\rm FIR} \rangle = 4.4\pm0.1 \, {\rm L}_{\odot}$ and
$\langle L_{\rm FIR} \rangle = 7.6\pm0.4 \, {\rm L}_{\odot}$;
$\langle L_{\rm FIR}/M \rangle = 1.0\pm0.1 \, {\rm L}_{\odot} {\rm M}_{\odot}^{-1}$ and
$\langle L_{\rm FIR}/M \rangle = 1.7\pm0.1 \, {\rm L}_{\odot} {\rm M}_{\odot}^{-1}$.
We note that these average values
are essentially equivalent to those shown in Table~\ref{tab:prevsproto} 
and thus do not change any of our previous conclusions.

We also previously mentioned in \S\ref{sec:separating} that outside of the region
covered by MIPS24 we used {\it MSX} to separate starless and proto-stellar cores.
It might be argued, then, that the trend we find in parameter space between
starless and proto-stellar cores may be caused partially by our inability to
detect faint MIR sources associated with BLAST cores using {\it MSX} data.
However, we think this is not the case for two reasons: (i) if some of the cores
outside the MIPS24 coverage are incorrectly classified as starless because
of the lack of a {\it MSX} counterpart, then the overall starless sample will be
contaminated by (likely) {\it warmer} proto-stellar cores, thus {\it reducing}
the difference in the median temperature between starless and
proto-stellar cores;  (ii) within the MIPS24 map, when we consider a more
complete sample ($M> 4.2 \, {\rm M}_\odot$, see Figure~\ref{fig:corefraction}) 
the trend in physical parameter space discussed above is
confirmed: the median temperature (or $L/M$ ratio) of starless and proto-stellar cores
are, respectively, 14.2 and 15.4~K (or 0.9 and 1.4~${\rm L}_\odot\,{\rm M}_\odot^{-1}$, respectively).
In addition, we note that 71\,\% of the BLAST cores associated with a {\it MSX} counterpart are
concentrated in the region covered by MIPS24 (measuring about $1.8\,$deg$^2$), 
while the remaining 29\,\% is found outside this area (which measures $\simeq 2.2\,$deg$^2$). Thus, 
using {\it MSX} data alone, we find a larger density of candidate proto-stellar cores toward the
central region of Vela-D, suggesting that this region is in a later stage of evolution. It also 
indicates that the number of BLAST cores with a MIR counterpart should be expected to decrease
outside the MIPS24 map, which in turn makes the lack of sensitive MIR observations outside the 
MIPS24 area less critical.  In fact, if we use the MSX source surface density as a rough indicator of the
density of proto-stellar sources (correcting for the area ratio, $1.8/2.2$, one expects a $\sim 25$\% 
probability of finding a proto-stellar core {\it outside} the MIPS24 area),
and the ratio of the total number of MIPS24 sources to the number of 
MIPS24 sources with a MSX counterpart in the MIPS24 area ($\simeq 5$), we thus estimate that
the number of proto-stellar cores {\it not} detected by MSX outside the MIPS24 area is $\sim 10$.

\subsection{Mass Spectrum}

\begin{figure}
\centering
\includegraphics[width=5cm,angle=270]{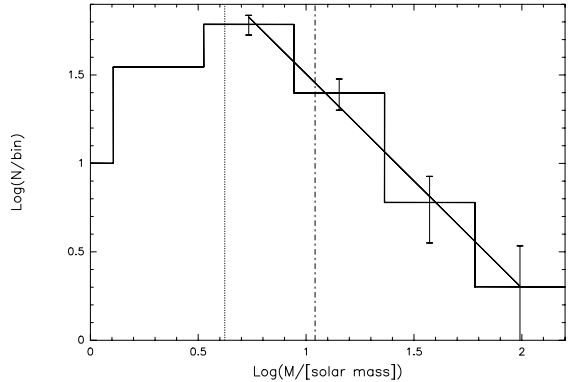}
\vspace{2mm}
\caption{
  Combined starless and proto-stellar core mass function of the 141
BLAST objects associated with Vela-D. Poisson error bars are shown;
however, additional sources of uncertainties are associated with the
parameters of Equation~(\ref{eq:mass}). The masses in this plot are calculated
assuming a dust emissivity index, $\beta = 2.0$, a dust mass absorption coefficient,
$\kappa_{0} = 16$\,cm$^2$\,g$^{-1}$ at 250\,\micron\ , a gas to dust ratio of 100 and a
distance of 700 pc.  The vertical dot-dashed line
represents the mass completeness limit for sources warmer than 10~K.
However, because almost all of the Vela-D sources have temperatures $> 12$~K
(\S\ref{sec:masslum})
we also show the completeness limit for cores warmer than 12~K
(dotted line).  The solid line
represents the best-fit to the CMF, yielding a power-law slope of
$\alpha = -2.2$. The lowest bins suggest a turnover, but this
shape is likely due to incompleteness.
Likewise, the bin at $M \simeq 5.6 \,$M$_{\odot}$ is also likely to be somewhat affected
by completeness.
  }
\label{fig:masspec}
\end{figure}

It is interesting to investigate the combined mass spectrum,
including both starless and proto-stellar sources in Vela-D.  The masses
of individual cores were placed in logarithmically spaced bins and
a lower limit on the error was estimated from
the Poisson uncertainty for each bin. The resulting mass function
is shown in Figure~\ref{fig:masspec}. We then fit a power-law
$(dN/dM \propto M^\alpha$) to the CMF for $M > 4 \, {\rm M}_{\odot}$, 
finding a slope $\alpha = -2.2$ with a correlation  coefficient $r = 0.99$.
The best-fit slope depends somewhat on the histogram binning. Given the
relatively low number of sources in Vela-D we could not vary greatly
the bin widths; however, for bin widths of $\simeq 1.7 $ to 4.2~${\rm M}_{\odot}$ we 
obtain values $\alpha \simeq -2.1$ to $\alpha \simeq -2.5$, so we empirically assign the
value $\alpha = -2.3 \pm 0.2$. We note that in the fit the bin at $M \simeq 5.6\,$M$_{\odot}$
has also been used, but because this bin is likely to be affected somewhat by 
completeness effects, it may be biasing {\it low} the resulting slope of the CMF.

%
%

Because almost all of the Vela-D sources have temperatures $> 12\,$K
(see \S\ref{sec:masslum})
the appropriate mass completeness limit is shown as a vertical
dotted line in Figure~\ref{fig:masspec} for cores warmer than 12~K.
The two lowest bins suggest a turnover, but this
shape is probably due to incompleteness and thus cannot be used to
make any statement about the point at which the mass distribution flattens.
Compared to previous surveys and measurements of the CMF, our data
rely on a robust determination of the core temperature. There may
be some concern over the use of a fixed value of the dust emissivity,
$\beta$; however, as discussed by \citet{chapin2008} this effect may shift 
the CMF, but does not affect the power-law index. 

The slope of the CMF found for the cores in Vela-D is very similar 
to that determined for Orion ($\alpha = -2.35$) 
by \citet{nutter2006} in
the mass range $M > 2.4 \, {\rm M}_\odot$, and also to that found
for the cores in Perseus, Serpens and Ophiucus ($\alpha = -2.3 \pm 0.4$)
by \citet{enoch2008},
in a comparatively lower mass range ($M > 0.8 \, {\rm M}_\odot$).  
Our CMF is steeper than that found also in Vela-D by \citet{massi07}, who report 
$\alpha \sim -1.9 \, {\rm to} \, -1.4$, and by \citet{elia2007} who find
$\alpha \sim -2.0 \, {\rm to} \, -1.3 $ (with $\alpha \sim -2$  as the most probable). 
The lower values for $\alpha$ measured by \citet{massi07} and \citet{elia2007}
are likely a consequence of the smaller source samples used to compute the CMF and also
the fact that these surveys may have missed many of
the coldest, low-mass sources. In fact, if we evaluate the CMF using only the 
BLAST cores with SIMBA counterparts we find values of $\alpha$ consistent with those
of \citet{massi07} and \citet{elia2007}.

All of these CMFs appear to have remarkably similar shapes to that of the
stellar IMF, as given for example by \citet{kroupa2002}. While the
observations taken together suggest that the CMF is a direct precursor of the stellar
IMF, it should be noted that the measured masses of \citet{nutter2006},
 \citet{enoch2008} and \citet{massi07} are somewhat more uncertain than ours because
they lack (partially or completely) temperature information for the cores.

We have not attempted to evaluate separately the CMF of starless and proto-stellar 
cores in Vela-D, because of the relatively small number of sources in the two
sub-samples. However, this has been done for the much larger source sample of 
the Vela deep map by \citet{netterfield2009}, who use this result to infer the
lifetime of the BLAST starless cores in the VMR. \citet{netterfield2009} 
find that cores above 14~M$_\odot$
in the Vela deep map follow a CMF with $\alpha = -2.77 \pm 0.16$, which is moderately steeper
than the CMFs mentioned earlier  and  also steeper than 
the Salpeter IMF. However, they also find that in the
smaller Vela-C region,  the CMF has $\alpha = -2.55 \pm 0.20$ thus consistent within the errors
with the value found in Vela-D, also because, as previously mentioned, 
we  may be biasing low the slope of the CMF. 
We note that the somewhat less steep CMF in Vela-D, which is in a 
later evolutionary phase compared to Vela-C, 
is consistent with cold cores following a steeper mass function 
than more evolved cores, as found by \citet{netterfield2009}.

\section{SUMMARY AND CONCLUSIONS}
\label{sec:concl}

In this paper we have presented a detailed analysis of the dense cores in the Vela-D
molecular cloud, utilizing the sensitive maps at 250, 350 and 500\,\micron\ 
 obtained by BLAST during its 2006 science flight from Antarctica.
We have combined the multi-color BLAST photometry with previous MIR, FIR, 
and millimeter-continuum observations to determine the physical parameters 
of the population of cores detected by BLAST.
Our sample includes a total of 141 objects, although only 89 cores 
are located within the common area covered by the MIPS24 and MIPS70  maps.
We summarize the other results of this paper as follows:

1. The multi-band photometry, combined with an isothermal modified blackbody model, allows
us to analyze the effects that the uncertainties on the derived physical parameters
of the individual sources have on the overall properties of starless and
proto-stellar cores.

2. In the area of Vela-D covered by the
MIPS24, MIPS70 and SIMBA maps we find 34 cores, out of 89, having 
no MIPS24 or MIPS70
counterpart, with 28 of these cores not having been detected by SIMBA either.
Therefore, these observations demonstrate the importance of observing the early cores
at or near the SED peak, as made possible by BLAST.

3. The CMF determined from all (starless and proto-stellar)
BLAST cores in Vela-D shows a power-law slope of $\alpha = -2.3 \pm 0.2$,
consistent, within the errors, with other continuum (sub)millimeter surveys and in particular with
the slope found by \citet{netterfield2009} in Vela-C.

4. We find a moderate difference between the median mass of proto-stellar and starless cores.
We also find that proto-stellar cores are somewhat warmer and more luminous than starless 
cores, and that there is a progressively higher fraction of proto-stellar cores with increasing
core temperature.  This suggests, as expected, a luminosity and temperature evolution due to 
the appearance of an embedded protostar.

5.  In terms of the observational properties, there appear to be a smooth transition from 
the pre- to the proto-stellar phase, as suggested by  differences found within
the two populations. In particular, for proto-stellar cores we find a correlation between 
the MIPS24 flux, associated with the central protostar, and the temperature of the dust
envelope. 
Our results can thus provide guidelines for understanding the physical conditions 
where the transition between pre- and proto-stellar cores takes place.

\acknowledgments
We acknowledge the support of NASA through grant numbers NAG5-12785,
NAG5-13301, and NNGO-6GI11G, the NSF Office of Polar Programs, the Canadian
Space Agency, the Natural Sciences and Engineering Research Council (NSERC) of
Canada, and the UK Science and Technology Facilities Council (STFC).
This work is also based, in part, on observations made with the {\it Spitzer
Space Telescope}, which is operated by the Jet Propulsion Laboratory,
California Institute of Technology under a contract with NASA. Support
for this work was provided by NASA through an award issued by JPL/Caltech.
L. O. acknowledges partial support by the Puerto Rico Space Grant Consortium and
by the Decanato de Estudios Graduados e Investigaci\'on of the University
of Puerto Rico. 

\bibliographystyle{apj}
\bibliography{apj-jour,refs}

\clearpage
\LongTables

\begin{landscape}

%
%
\begin{deluxetable}{lcccccccccc}
\tablewidth{0pt}
\tabletypesize{\scriptsize}
\tablecaption{BLAST Sources \label{tab:src}}
\tablehead{
%
\colhead{Source \#} &
\colhead{Source name} &
\colhead{$l$} &
\colhead{$b$} &
\colhead{$S_{250}$} &
\colhead{$S_{350}$} &
\colhead{$S_{500}$} &
\colhead{$T$} & 
\colhead{$M$} & 
\colhead{$L$} & 
\colhead{Deconvolved} \\
%
\colhead{} &
\colhead{} &
\colhead{} &
\colhead{} &
\colhead{} &
\colhead{} &
\colhead{} &
\colhead{} &
\colhead{} &
\colhead{} &
\colhead{diameter} \\
%
\colhead{} &
\colhead{} &
\colhead{deg} &
\colhead{deg} &
\colhead{(Jy)} &
\colhead{(Jy)} &
\colhead{(Jy)} &
\colhead{(K)} &
\colhead{(${\rm M}_\odot$)} &
\colhead{(${\rm L}_\odot$)} &
\colhead{(pc)} 
}
\startdata
%
%
%
   0 & BLAST J084441-431144 & 263.0044 & $-0.2518$ & $  7.2\pm 0.9$ & $  4.7\pm 0.7$ & $  2.5\pm 0.5$ & $ 12.9\pm^{  1.3}_{  0.9}$ & $  3.5\pm^{  1.2}_{  1.1}$ & $   1.7\pm^{  0.5}_{  0.3}$ & 0.14 \\
   1 & BLAST J084508-433755 & 263.3975 & $-0.4588$ & $  6.5\pm 0.9$ & $  4.5\pm 0.8$ & $  2.0\pm 0.6$ & $ 15.7\pm^{  0.4}_{  0.3}$ & $  1.5\pm^{  0.2}_{  0.4}$ & $   2.4\pm^{  0.3}_{  0.3}$ & 0.14 \\
   2 & BLAST J084509-434544 & 263.5011 & $-0.5379$ & $  4.6\pm 0.7$ & $  3.4\pm 0.6$ & $  1.6\pm 0.4$ & $ 12.3\pm^{  1.7}_{  1.1}$ & $  2.8\pm^{  1.3}_{  1.1}$ & $   1.1\pm^{  0.4}_{  0.2}$ & 0.10 \\
   3 & BLAST J084531-435006 & 263.6001 & $-0.5302$ & $210.1\pm28.1$ & $ 93.3\pm19.6$ & $ 30.5\pm13.3$ & $ 20.9\pm^{  0.4}_{  0.4}$ & $ 23.2\pm^{  3.2}_{  2.5}$ & $ 206.5\pm^{ 11.9}_{ 14.8}$ & 0.17 \\
   4 & BLAST J084535-435156 & 263.6303 & $-0.5414$ & $271.9\pm30.4$ & $117.7\pm18.3$ & $ 43.2\pm10.5$ & $ 20.9\pm^{  0.4}_{  0.4}$ & $ 28.5\pm^{  3.8}_{  3.1}$ & $ 253.5\pm^{ 14.9}_{ 18.6}$ & 0.18 \\
   5 & BLAST J084538-435507 & 263.6770 & $-0.5678$ & $ 13.2\pm 1.6$ & $  7.9\pm 1.2$ & $  4.1\pm 0.8$ & $ 13.8\pm^{  1.3}_{  0.9}$ & $  4.7\pm^{  1.5}_{  1.5}$ & $   3.5\pm^{  1.0}_{  0.7}$ & 0.12 \\
   6 & BLAST J084539-435133 & 263.6334 & $-0.5273$ & $281.5\pm40.9$ & $134.8\pm29.5$ & $ 48.2\pm19.9$ & $ 16.2\pm^{  0.2}_{  0.2}$ & $ 55.9\pm^{  6.8}_{  5.3}$ & $ 106.5\pm^{  5.9}_{  7.0}$ & 0.21 \\
   7 & BLAST J084540-430400 & 263.0150 & $-0.0316$ & $  8.0\pm 1.4$ & $  6.6\pm 1.2$ & $  3.7\pm 0.8$ & $ 11.1\pm^{  1.6}_{  1.1}$ & $  8.2\pm^{  4.7}_{  3.2}$ & $   1.6\pm^{  0.5}_{  0.4}$ & 0.15 \\
\enddata
\tablecomments{Flux densities for BLAST sources are quoted at
  precisely 250, 350, and 500\,\micron\ using SED fits to obtain
  color-corrections for the band-averaged flux densities \citep{chapin2008, truch2008}.
  The quoted statistical uncertainties are determined from Monte Carlo simulations
  and do not include calibration uncertainties (see \S\ref{sec:sed}).
The deconvolved or intrinsic source FWHM is obtained from 
the of best-fit source size, $\theta_{\rm fit}$, and the BLAST beam size, $\theta_{\rm b}$, as
$\theta_{\rm dec} = (\theta_{\rm fit}^2 - \theta_{\rm b}^2)^{1/2}$ \citep{netterfield2009}.  
}

\end{deluxetable}
\clearpage
\end{landscape}

%
%
\begin{deluxetable}{lcccccccccccc}
\tablewidth{0pt}
\tabletypesize{\scriptsize}
\tablecaption{{\rm 1.2mm-24\,\micron}  counterparts \label{tab:flux}}
\tablehead{
\colhead{Source \#}   &
\colhead{Source name}       &
\colhead{$S_{1200}$} &
\colhead{$N_{1200}$}   &
\colhead{$S_{100}$}  &
\colhead{$S_{60}$}   &
\colhead{$N_{\rm IRAS}$}   &
\colhead{$N_{\rm Akari}$}  &
\colhead{$S_{70}$}   &
\colhead{$N_{70}$}     &
\colhead{$S_{24}$}   &
\colhead{$N_{24}$}     &
\colhead{$N_{\rm MSX}$}     \\
\colhead{}     &
\colhead{}     &
\colhead{(Jy)} &
\colhead{}     &
\colhead{(Jy)} &
\colhead{(Jy)} &
\colhead{}     &
\colhead{}     &
\colhead{(Jy)} &
\colhead{}     &
\colhead{(Jy)} &
\colhead{}     &
\colhead{}     \\
}
\startdata
%
%
   0 & BLAST J084441-431144 & $   -^{ }$ & -- & $    3.1^{a}$ & $   0.1^{a}$ &  0 &  0 & $      -^{ }$ & -- & $     -^{ }$ & -- &  0 \\
   1 & BLAST J084508-433755 & $0.40^{ }$ &  0 & $   12.6^{a}$ & $   1.5^{a}$ &  0 &  1 & $   0.31^{ }$ &  1 & $ 0.017^{ }$ &  1 &  0 \\
   2 & BLAST J084509-434544 & $0.40^{ }$ &  0 & $      -^{ }$ & $     -^{ }$ &  0 &  0 & $      -^{ }$ &  0 & $     -^{ }$ &  0 &  0 \\
   3 & BLAST J084531-435006 & $3.17^{ }$ &  1 & $ 1688.2^{a}$ & $ 620.6^{a}$ &  0 &  0 & $ 131.24^{a}$ &  1 & $ 1.934^{ }$ &  1 &  1 \\
   4 & BLAST J084535-435156 & $3.70^{ }$ &  1 & $ 1573.0^{a}$ & $ 592.0^{a}$ &  0 &  2 & $ 161.44^{a}$ &  1 & $ 1.280^{ }$ &  2 &  0 \\
   5 & BLAST J084538-435507 & $0.40^{ }$ &  0 & $  123.8^{a}$ & $   1.3^{a}$ &  0 &  0 & $   5.77^{a}$ &  0 & $ 0.194^{ }$ &  1 &  0 \\
   6 & BLAST J084539-435133 & $3.59^{ }$ &  1 & $ 1519.9^{a}$ & $ 685.5^{a}$ &  0 &  0 & $  16.97^{ }$ &  1 & $     -^{ }$ &  0 &  1 \\
   7 & BLAST J084540-430400 & $   -^{ }$ & -- & $      -^{ }$ & $     -^{ }$ &  0 &  0 & $      -^{ }$ & -- & $     -^{ }$ & -- &  0 \\
\enddata
\tablecomments{Flux density ($S$) and number ($N$) of sources associated
  with BLAST objects for each catalog considered: SIMBA, {\it IRAS}, {\it Akari}, MIPS70, MIPS24 and {\it MSX}. 
Fluxes are catalog fluxes, unless noted otherwise.
When $N=0$, but the flux density $S$ is not zero, then an upper limit to the flux is listed, either 
estimated by aperture photometry (indicated by ``a'') or by determining the noise in the map.
The search radius is a variable function of both
  BLAST and {\it IRAS} positional uncertainties (see
  \S\ref{sec:blastflux}).}
\tablenotetext{a}{Aperture photometry has been used.}
\end{deluxetable}

\end{document}